\documentclass[Phys,submission]{SciPost}

\makeatletter
\def\@efAtBegDocument#1{}   
\makeatother

\usepackage{graphicx}
\usepackage{dcolumn}
\usepackage{bm}
\usepackage{amsmath,amssymb}
\usepackage[caption=false]{subfig}
\usepackage{xcolor}
\usepackage{comment}

\hypersetup{
    colorlinks=true,
    linkcolor=blue,
    citecolor=blue,
    urlcolor=blue
}

\renewcommand{\subsectionmark}[1]{}

\renewcommand{\vec}[1]{\boldsymbol{#1}}

\renewcommand{\(}{\left(}
\renewcommand{\)}{\right)}
\renewcommand{\v}[1]{{\bf #1}}

\begin{document}

\begin{center}
{\Large \textbf{Spin-incoherent Mott-Thouless pumps}}
\end{center}

\begin{center}
Ajesh Kumar$^{1*}$, Urban F. P. Seifert$^1$, Erez Berg$^{2,3,4}$, and Achim Rosch$^1$
\end{center}

\begin{center}
{\small
{\bf 1} Institute for Theoretical Physics, University of Cologne, 50937 Cologne, Germany \\
{\bf 2} Department of Condensed Matter Physics, Weizmann Institute of Science, Rehovot 7610001, Israel \\
{\bf 3} Materials Department, University of California, Santa Barbara, CA 93106, USA \\
{\bf 4} Department of Electrical and Computer Engineering, University of California, Santa Barbara, CA 93106, USA \\[1ex]
* \href{mailto:ajesh.kumar@uni-koeln.de}{ajesh.kumar@uni-koeln.de}
}
\end{center}

\begin{center}
\today
\end{center}

\section*{Abstract}
{\bf
A Thouless pump describes a system in which a quantized amount of charge is transported by one lattice spacing per cycle when the parameters of a Hamiltonian are varied slowly and periodically.
In the standard case, this quantization requires the system to remain in its ground state with vanishing thermodynamic entropy throughout the pumping process. Here, we introduce a class of Mott-Thouless pumps, which operate in highly entropic, spin-incoherent Mott states. We show analytically that these states exhibit exponentially protected quantized transport despite their extensive entropy, because the pumped charge resides in a gapped sector that remains dynamically decoupled from the hot spin degrees of freedom. By contrast, motivated by a recent experimental realization, we identify other classes of Mott–Thouless pumps that lack this protection. In these systems, spin excitations can efficiently generate charge excitations, leading to a rapid breakdown of quantized transport. Our analytical results are supported by numerically exact real-time simulations of finite systems.
}

\vspace{10pt}
\noindent\rule{\textwidth}{1pt}
\tableofcontents
\noindent\rule{\textwidth}{1pt}
\vspace{10pt}

\section{Introduction}
Quantized transport is a hallmark of topological phenomena in condensed matter systems. A paradigmatic example is the Thouless pump \cite{Thouless1983}, where adiabatic and periodic modulation of a Hamiltonian’s parameters leads to the transfer of an integer charge per cycle. From a topological viewpoint \cite{Citro2023}, one-dimensional Thouless pumps are closely related to two-dimensional Chern insulators. In this picture, the adiabatic time evolution acts as an additional dimension, so that the system explores a two-dimensional parameter space spanned by crystal momentum and time. The pumped charge per cycle is then determined by the corresponding Chern number.

While such pumps can be realized with non-interacting fermions \cite{Thouless1983}, it was recognized early on that they remain robust against interactions and disorder \cite{NiuThouless1984}, owing to a finite energy gap separating the ground state from the excitations.
In certain cases -- such as bosonic systems \cite{Berg2011} -- interactions are even required to open the gap, and interacting Thouless pumps have  been studied intensively in recent years \cite{NiuThouless1984,Shindou2005,Berg2011,Ohyama2022,Citro2023,ArguelloLuengo2024,Walter2023,Viebahn2024}.
Experimental realizations of Thouless pumps have been demonstrated in a variety of physical platforms, including ultracold atomic gases in optical lattices \cite{Lohse2016ThoulessPump,Nakajima2016TopologicalPump,Schweizer2016SpinPump,Viebahn2024,Walter2023,Viebahn2024}, photonic systems \cite{Kraus2012QuasicrystalPump,Verbin2013PhotonicQuasicrystal,Juergensen2021}, and mechanical metamaterials \cite{Grinberg2020MagnetoMechanical}.
Beyond their role in model systems and in understanding topological phases, quantized pumps also provide a conceptual foundation for quantum metrology. For example, a sufficiently fast  quantized pump can be used to realize the ampere with high precision \cite{Ampere2016}.


While the concept of a Thouless pump has originally been formulated in terms of adiabatic evolution of quantum ground states and the associated geometric transport in coherent systems, it is important to understand how pumping mechanisms are modified in the presence of decoherence and dissipation. This question has been intensively studied for mesoscopic pumps \cite{Zhou1999MesoscopicPump}.
It has, for example, been shown that parametric pumping can persist even in the presence of inelastic processes, where the pumping mechanism  crosses over from coherent interference–based transport to a purely classical mechanism \cite{Moskalets2001InelasticPump,Moskalets2002NoisePump,Cremers2002DephasingPump,Avron2002OpenPump,Rahav2008StochasticPumps}. 

Similarly, the question arises of how finite temperatures and an external bath affect  pumping in one-dimensional systems. A bosonic bath can, for example, destroy quantized pumping \cite{HeidrichMeisner2022} or, under the right conditions, enhance it \cite{Arceci_2020}. There has also been significant interest in defining topological invariants not only for pure ground states but also for mixed states or density matrices. Bardyn \emph{et al.} demonstrated that for mixed fermionic states, e.g., at finite temperature, an ``ensemble geometric phase'' (EGP) can be defined, which remains quantized when measured through appropriate many-body observables \cite{Bardyn2018}. Subsequent work has extended the classification of symmetry-protected topological phases to mixed states \cite{ma2025,guo2025,xue2024tensornetworkformulationsymmetry} and explored the topology of Gaussian mixed states using full counting statistics \cite{fleischhauer2021}.

\begin{figure}
    \centering
    \includegraphics[width=0.6\linewidth]{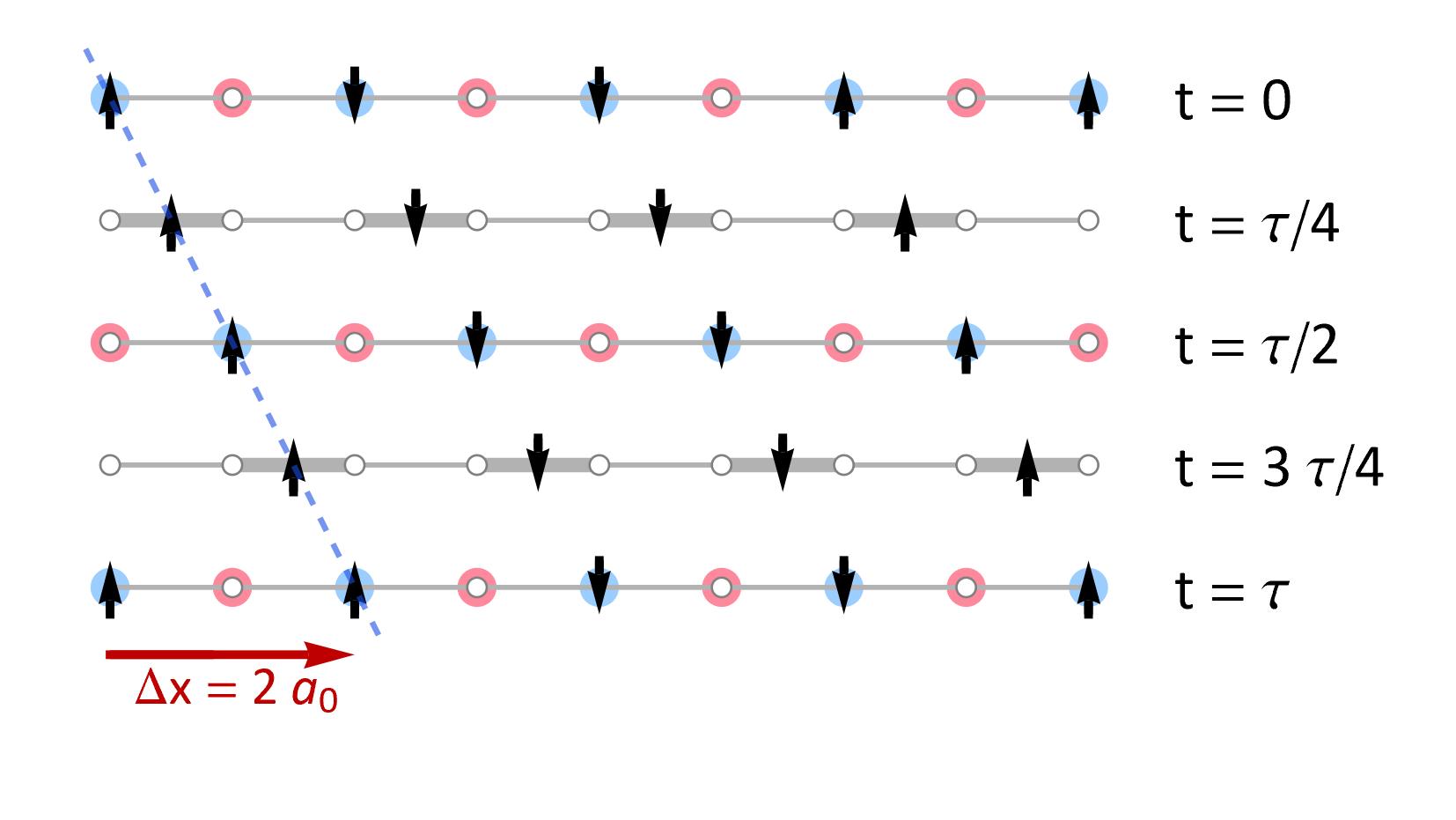}
\caption{Schematic illustration of a quarter-filled Mott-Thouless pump  described by $H_{\rm RM}$, Eq.~\eqref{eq:H1}, with one fermion per two-site unit cell. Pumping is driven by a staggered potential (red and blue circles) and a staggered hopping amplitude (thick and thin gray bonds). During one pumping cycle, a unit charge is transported by one unit cell, $\Delta x = 2a_0$. In the  limit of strong repulsive interactions, the charge dynamics effectively decouples from the spin dynamics, so that quantized charge transport remains robust even in highly spin-incoherent states with large spin entropies.}
    \label{fig:pump1_schematic}
\end{figure}
Here, in contrast, we focus on the dynamical robustness of quantized transport in highly entropic many-body states. We investigate 
a class of ``Mott-Thouless pumps'', where the system is initialized not in a pure ground state but in a spin-incoherent Mott state with entropy approaching $\ln 2$ per particle, see Fig.~\ref{fig:pump1_schematic}. This is motivated both by the  question of whether exponentially protected pumping is possible very far from the ground state and by the observation that -- despite remarkable recent progress \cite{Xu2025} -- it remains a major challenge to reach low-entropy states in fermionic ultracold atom experiments. 
We show that quantized Thouless pumping can remain exponentially protected even in high-entropy spin-incoherent Mott states for suitable classes of Mott-Thouless pumps. By exponentially protected we mean that deviations from quantized transport and the rates for creating detrimental charge excitations are exponentially suppressed in the relevant separation of energy scales.

The key reason is that the pumped charge resides in a gapped sector that  couples only through exponentially suppressed processes to the hot spin degrees of freedom.
Our approach relaxes the traditional requirement of ground-state initialization in Thouless pumps. More generally, our results show that exponentially protected topological transport can survive even in highly entropic many-body states  if the entropy resides in sectors that remain effectively decoupled. The exponentially slow equilibration between spin and charge sectors connects to the broader phenomenology of weakly broken conservation laws, prethermalization, and generalized Gibbs ensembles in driven many-body systems \cite{Abanin2017,Lange2017,Lange2018}.

Strongly interacting one-dimensional electron systems can  successfully be described within the Luttinger liquid framework, where spin and charge excitations are decoupled \cite{Fiete07}. However, when the spin excitation energy $J$ is much smaller than the thermal energy $ T \gg J$, the system enters a universal ``spin-incoherent'' regime \cite{Matveev04,Balents04,Fiete07,FeiguinFiete2010}. In this regime, charge dynamics and charge correlations remain  well ordered, while the spin sector is highly excited and incoherent. This spin-incoherent regime is particularly relevant for strongly correlated, low-density, or multi-component systems with small exchange couplings $J$ and is the basis for our analysis.

The remainder of the paper is organized as follows. In Sec.~\ref{sec:models}, we introduce the models and discuss exactly solvable limits. In Sec.~\ref{heatingQuarter}, we analyze heating and the stability of quantized pumping. In Sec.~\ref{sec:half}, we discuss a recent experimental realization of a Mott-Thouless pump \cite{Walter2023,Viebahn2024} and its lack of exponential protection, before concluding the paper.

\section{Two models with exactly solvable limits}\label{sec:models}
We will mainly consider two types of Mott-Thouless pumps using either a lattice model or a continuum model as the starting point.
We first consider an interacting variant of the Rice-Mele model \cite{RiceMele82}, originally used to model diatomic polymers
\begin{align}\label{eq:H1}
H_{\text{RM}}(t)=& -\sum_{i,\sigma=\uparrow/\downarrow} (t_0+\delta t(t)\, (-1)^i)c^\dagger_{i,\sigma} 
c_{i+1,\sigma}+ h.c. \nonumber \\
&+ \delta v(t)\,
\sum_{i,\sigma=\uparrow/\downarrow} (-1)^i n_{i,\sigma}+U \sum_i n_{i\uparrow}n_{i\downarrow},
\end{align}
where we denote the lattice constant by $a_0$. The unit-cell spacing is $a=2 a_0$. This model has been realized with high precision using ultracold atoms~\cite{Walter2023,Viebahn2024} in the strong-interaction limit. 

The staggered hopping $\delta t$ and the staggered potential $\delta v$ are assumed to change slowly in time 
\begin{align}\label{path1}
\delta t(t)=\delta t \sin(\omega t)
, \qquad \delta v(t)=\delta v \cos(\omega t).
\end{align}
Throughout the paper, we assume the drive to be adiabatic with respect to the charge gap, but not necessarily with respect to the spin dynamics.

For $U=0$, the corresponding single-particle Hamiltonian $H^{(1)}_k(t)$ 
can be written in momentum space as 
\begin{align}
H^{(1)}_k(t)&=\vec n_k  \cdot \vec \sigma \\
n^x_k&= \delta v(t), \ n^y_k=2 \delta t(t) \sin a_0 k, \  n^z_k=-2 t_0 \cos a_0 k,
\nonumber 
\end{align}
where we use states with momentum $k$ and $k+\pi/a_0$ as a basis,  $-\frac{\pi}{2a_0}< k \le \frac{\pi}{2a_0}$.
Using instantaneous eigenstates
$|\pm,k,t\rangle$, with  $H(t)|\pm,k,t\rangle=\epsilon_n(k,t)|n,k,t\rangle$ and $\epsilon_n(k,t)=\pm |\vec n_k|$, one obtains in the  two-dimensional space spanned by $k$ and $t$ two Chern bands with a minimal gap set by $\Delta_W=\min(4 \delta t,2 \delta v)$ and a
winding number $\pm 1$.
This non-interacting system is at temperature $T=0$ a Thouless pump \cite{Thouless1983} at half-filling, which transports 2 electrons (one for each spin direction) per cycle by  $a=2a_0$, the size of the unit cell in our system. Below, we will consider instead the limit of quarter filling, see Fig.~\ref{fig:pump1_schematic} for illustration.

For $U=\infty$, one can use a well-known trick \cite{spinless1990} to solve the model. One uses the fact that for $U=\infty$ there is never a spin-up and a spin-down particle on the same site. This constraint can be implemented simply by using {\em spinless} fermions, one for each spinful particle, where Pauli's exclusion principle guarantees that two particles are never on the same site. These spinless fermions fully capture the charge dynamics of the system for $U=\infty$. In contrast, the spin degrees of freedom can take random values without affecting the charge dynamics in the $U=\infty$ limit.

If we consider a system with quarter filling, i.e., one electron per unit-cell (consisting of two sites), we obtain an insulating state with a charge gap $\Delta_W$.  The spin excitations do not acquire dynamics  for $U=\infty$ and are described by the spin Hamiltonian $H_s=0$. The charge pumping can be computed from the non-interacting spinless-fermion model: per cycle one particle is transported by the distance $a=2a_0$, which is half of the pumped charge obtained in the half-filled non-interacting system.

The exact solution sketched above proves the existence of a spin-incoherent Mott-Thouless pump at $U=\infty$.
Before discussing what will happen at large but finite $U$, we briefly discuss another `trivial' example of an exactly solvable Mott-Thouless pump. Here, we consider a continuum model of spinful fermions in a time-dependent potential,
\begin{align}\label{eq:H2}
H_{sp}=&\sum_{\sigma=\uparrow,\downarrow}
\int dx \Biggl(\frac{\hbar^2  \nabla \Psi_\sigma^\dagger(x) \nabla \Psi_\sigma(x)}{2 m}+g n_\uparrow(x) n_\downarrow(x)\nonumber \\
&+\left(V_0(t) \cos(k_0 x) + V_1(t) \sin(k_0 x) \right)\Psi^\dagger_\sigma (x) \Psi_\sigma(x)\Biggr)
\end{align}
with $n_{\sigma}=\Psi^\dagger_{\sigma}\Psi_{\sigma}$. For $V_0(t)=V_0 \cos(\omega t)$ and $V_1(t)=V_1 \sin(\omega t)$, the total potential is given by \begin{align}
    V(x,t)=\frac{V_0+V_1}{2} \cos(k_0 x - \omega t)+\frac{V_0-V_1}{2} \cos(k_0 x + \omega t).
\end{align} 
For $V_0=V_1$ the system realizes a sliding potential and one can simply use a Galilean transformation, $x \to x-v_0 t$ with $v_0=\omega/k_0$ to map this problem to an equilibrium problem with a static potential $V_0 \cos(k_0 x)$. For $V_0$ large compared to the recoil energy, $V_0\gg E_R$ with $E_R=\frac{(2 \hbar  k_0)^2}{2 m}$, the system maps to a Hubbard model with the lattice constant $a=2 \pi/k_0$, an exponentially small hopping rate $t_0$ and an effective interaction $U \propto g \gg t_0$. Thus, for half-filling, i.e., one particle per lattice site and for temperature $ T \ll U $ but $T\gtrsim t_0^2/U$, and thus comparable or larger than the exchange energy of localized spins, the system in the comoving frame of reference is in a spin-incoherent Mott-insulating state.
Transforming back to the laboratory frame, the  Mott state therefore corresponds to a spin-incoherent Thouless pump. It pumps one particle by the distance $a$ during one cycle of duration $\tau=2 \pi/\omega$, exactly as in the previous example. In this type of pump, the Mott gap $ U$ sets the size of the  charge gap, which is smaller than the band gap $U \ll \Delta_W$ in this setting. 

The absence of heating for $V_0=V_1$ arises from Galilean invariance and the precise form of the chosen potential. If one breaks Galilean invariance or chooses $V_0= V_1+\Delta V$ with $\Delta V \ll V_0$
the system will slowly heat up as we  discuss  in Sec.~\ref{sec:heating_continuum}.


\section{Heating of the driven system at quarter filling}\label{heatingQuarter}
As a next step, we want to analyze how deviations from exactly solvable idealized settings affect the operation of the Thouless pump. What happens if $U$ in $H_\text{RM}$, Eq.~\eqref{eq:H1} is not infinite but large? 
Similarly, the question arises how a breaking of Galilean invariance in $H_{sp}$, Eq.~\eqref{eq:H2}, affects the system for $0<|V_1-V_0|\ll V_0, V_1$.
As energy is not conserved in the periodically driven system, it is clear that the system ultimately will heat up to an infinite temperature state in the absence of some cooling mechanism. But how long will this take and what happens when the Thouless pump couples to reservoirs?

Our goal is to derive scaling arguments which can be used to explore how central parameters of the theory and how the proximity to  the exactly solvable limits affects the operation of generalized Thouless pumps in a spin-incoherent regime.
\begin{figure}[t]\centering
\includegraphics[width = 0.6 \columnwidth]{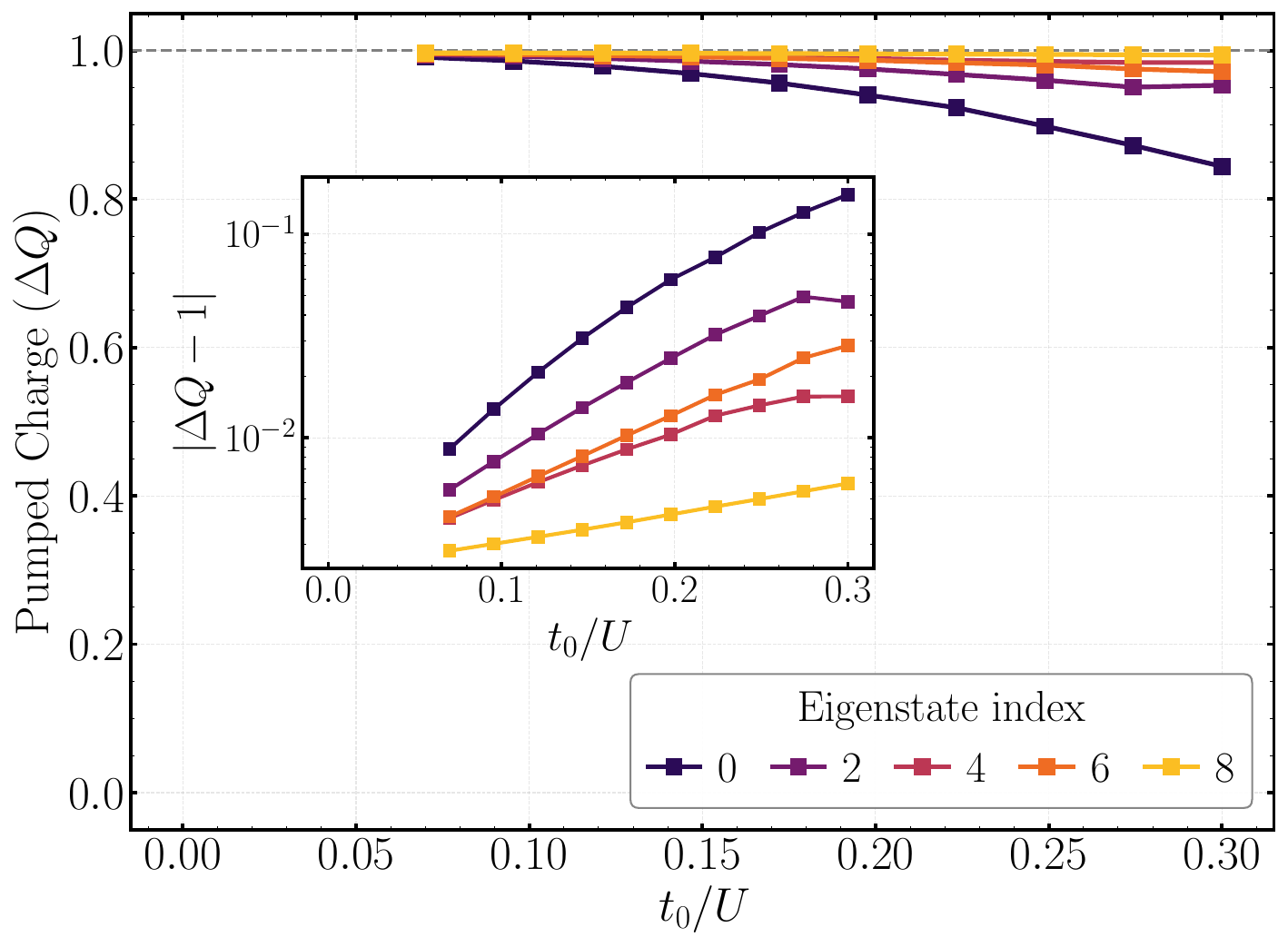}
\caption{Pumped charge per cycle as a function of $t_0/U$ for the quarter-filled Mott-Thouless pump described by Eq.~\eqref{eq:H1}. We show the time evolution for four low-energy initial states in the zero-momentum sector describing spin excitations. The curves are labeled by the energy-ordering index of the initial state, in ascending order.
Parameters: system size $L=16$, particle numbers $N_{\uparrow}=N_{\downarrow}=4$, $\delta t=0.5$, $\delta v=1$, and driving frequency $\omega=0.1$. Each initial state is evolved for five driving periods, and the transported charge is averaged over these cycles. The inset shows the deviation from perfect quantization on a logarithmic scale. In the $U \rightarrow \infty$ limit, the curves are expected to approach the same value, which is non-zero here due to finite-size effects. To ensure an adiabatic protocol, we use a frequency ramp in which the drive frequency is smoothly increased from $0$ to $\omega$ over a time $\tau$, see App.~\ref{app:numerics} for details and further numerical results.}
\label{fig:pump_vs_U}
\end{figure}

As an illustration, Fig.~\ref{fig:pump_vs_U} shows the pumped charge of the Mott-Thouless pump described by Eq.~\eqref{eq:H1} as a function of $t_0/U$ obtained from a numerical solution of a small system with $L=16$ sites, see App.~\ref{app:numerics} for details and also further results on the frequency dependencies. The different curves correspond to different initial states in the spin sector, chosen from low-lying eigenstates of the initial Hamiltonian. The quantization of the pumped charge improves systematically with increasing $U$, i.e., upon approaching the Mott limit. Importantly, spin excitations do not degrade the operation of the Thouless pump in the large-$U$ regime, on the contrary, the pumped charge approaches unity more closely for higher-energy spin states.

In the following, we focus on much larger systems, where the spin-excitation spectrum becomes dense. In this regime the maximal spin-excitation energy is of order $N J$, 
with $N$ the number of unit cells, $L/a$,
and $J$ the effective spin-exchange coupling, therefore exceeding the charge gap.

\subsection{Heating of an infinite system}
To analyze heating, we will first study the low-energy spin sector, adding the coupling to high-energy charge excitations in a second step. The effective Hamiltonian of the spins in a Mott state is given by the Heisenberg model
\begin{align}\label{eq:Heisenberg}
H_s(t)= \sum_{i,\delta} J_{\delta}(t)\,\vec S_{i+\delta} \cdot \vec S_i
\end{align}
with time-dependent exchange couplings $J_{\delta}(t)$.
Note that at quarter filling, there is one spin per two-site unit cell, and $J_{\delta}(t)$ is independent of $i$, consistent with translational invariance.

For $\delta t \sim \delta v \sim t_0$, we estimate the nearest-neighbor coupling $J_1 \sim t_0^2/U$ while longer ranged couplings are suppressed, $J_2 \sim t_0^3/U^2$, see App.~\ref{app:effective_Heisenberg} for details. 
At the same order in the strong-coupling expansion, three-spin 
terms of the form
$\vec S_i \cdot (\vec S_{i+1} \times \vec S_{i+2})$
are, in principle, also generated. Since these terms are odd under time reversal,
they vanish in the static limit and, for slowly varying drives, acquire an
additional suppression by a factor $\omega/\Delta_W$.

For strong correlations and one electron per unit cell, $H_s$ describes the spin sector for both models discussed above, Eq.~\eqref{eq:H1} and  Eq.~\eqref{eq:H2} (half filling of the lowest band) but the precise type of time-dependencies of the coupling constants $J_\delta$ is model specific.
It is important to stress that the index $i$ in Eq.~\eqref{eq:Heisenberg} does not describe a spin at a fixed position but instead per cycle of the Thouless pump a given spin moves by one lattice constant $a$, see Fig.~\ref{fig:pump1_schematic} and App.~\ref{app:effective_Heisenberg}.

Generically, the coupling terms $J_1(t)\gg  J_2(t) \gg J_3(t)\dots$ will depend periodically on time (the Galilean invariant case, $V_0=V_1$ in  Eq.~\eqref{eq:H2} being the only exception). If $[H_s(t),H_s(t')]\neq 0$ for $t\neq t'$ the spin sector will heat up. As \begin{align}
    [H(t),H(t')]\propto J_1(t) J_2(t')-J_2(t) J_1(t').
\end{align}
one obtains the dominant heating processes from the interplay of nearest and next-nearest coupling terms.

A useful trick is to absorb the time-dependence of the largest term $J_1$ in a rescaling of the time axis, $t \to \tilde t(t)$, using the ansatz $|\Psi(\tilde t(t))\rangle$ for the wavefunction with
\[\frac{d}{d t} |\Psi(\tilde t(t))\rangle=\frac{d\tilde t}{dt} \frac{d}{d \tilde t}  |\Psi(\tilde t(t))\rangle, \quad \frac{d\tilde t}{dt}=\frac{J_1(t)}{\bar J_1},\]
where $\bar J_1=\frac{1}{\tau} \int_0^\tau J_1(t)\, dt$ is the time-averaged coupling and $\tau=2 \pi/\omega$ is the period of the pump. This convention ensures that $\tilde t(n \tau)=n \tau$ and eliminates the time-dependence of $J_1$.

  With this rescaled time coordinate,
we can approximate $H_s$  as
\begin{align}
H_s(\tilde t) &\approx H_s^0+\Delta H(\tilde t) \\
H_s^0&= \bar J_1 \sum \vec S_{i+1} \cdot \vec S_i+\bar J_2 \sum \vec S_{i+2}\cdot \vec S_i, \nonumber 
\\ \Delta H(\tilde t)& \approx \delta J \cos(\omega \tilde t) \sum \vec S_{i+2}\cdot \vec S_i,
\nonumber \end{align}
where for simplicity we kept only the leading oscillating contribution with $\delta J\ll \bar J_1$. App.~\ref{app:effective_Heisenberg} provides a microscopic calculation of these terms. The term $\bar J_2$ ensures that integrability is broken in the static case, $\Delta H=0$, see Ref.~\cite{Lange2017,Lange2018} for a theory of driven approximately integrable systems. 
In App.~\ref{app:effective_Heisenberg} we argue that the next-nearest neighbor coupling $\delta J$ scales with $1/U^2$ in the large $U$ limit with $\bar J_1 \sim 1/U$. 
As an example, for $t_0=1$, $\delta t = 0.3$, $\delta v = 4$, we find that $\bar{J_2} \approx 0.016/U^2$ and $\delta J \approx -0.015/U^2$. 
 
Due to the oscillating term, the energy of our system increases. Using perturbation theory in $\delta J$, the increase of energy (per site and averaged over a time interval) can be computed from the imaginary part of the retarded susceptibility of the operator $\sum_i \vec S_{i+2} \cdot \vec S_i$,
\begin{align}
\frac{1}{N}\overline{\left\langle \frac{d}{d  t} H_s \right\rangle}=\frac{1}{2} \omega\, \text{Im} \chi(\omega) (\delta J)^2.\label{eq:heat1}
\end{align}
Here, $\chi(\omega)$ is computed relative to a thermal state, which is justified as long as the equilibration rate of $H_s$ is shorter than the relevant heating rates \cite{Genske15,Matthies22}.

Our goal is to obtain a qualitative estimate for the heating rates. For simplicity, we assume that the initial state has a high entropy and a large temperature $T\gtrsim \bar J$. Thus, we can use a high-temperature expansion of the system from which one obtains
$\text{Re} \chi(\omega=0)\sim \frac{1}{T}$. The real part is related to the imaginary part by a Kramers-Kronig relation, 
$\text{Re} \chi(0)=\int \frac{\text{Im }\chi(\omega)}{\omega}\, d\omega$. Here,  $\text{Im} \chi(\omega)$ is an odd function in $\omega$ with a width which is set by $\bar J_1$.

This constrains the possible form of $\text{Im} \chi(\omega)$.
Assuming a smooth spectral function with bandwidth set by $\bar J_1$, the oddness of $\text{Im} \chi(\omega)$ in $\omega$ then implies for small $\omega$
\begin{align}
\text{Im} \chi(\omega) \sim \frac{1}{T}\frac{\omega}{\bar J_1}.
\end{align}
Furthermore, the relation of average energy per site $e$ and temperature is obtained from the high-temperature expansion,
\begin{align}
e \sim -\frac{\bar J_1^2}{T}.\label{eq:T}
\end{align}
Thus, using Eq.~\eqref{eq:heat1}, we find
\begin{align}\label{eq:heatingE}
\partial_t e \sim -e\, \omega^2 \frac{(\delta J)^2}{\bar J_1^3}
\end{align}
or, equivalently,
\begin{align}
\frac{d T}{d t} \approx \frac{d T}{d e}\frac{d e}{dt} \sim  \omega^2\, T \,\frac{(\delta J)^2 }{\bar J_1^3}. 
\end{align}
Thus, the temperature rises exponentially for $t\to \infty$, but the prefactor becomes very small in the adiabatic limit $\omega \to 0$,
\begin{align}
T(t) \sim T_0 \exp\!\left({\alpha  \, \omega^2\, t \,\frac{(\delta J)^2 }{\bar J_1^3}}\right)
\end{align}
where $\alpha$ is a numerical factor of order $1$ which cannot be determined from our scaling analysis. 
From App.~\ref{app:effective_Heisenberg}, $\bar J_1 \sim \frac{t_0^2}{U}$ while $\delta J\sim \frac{t_0^3}{U^2}$. 
Thus, we expect that the time scale for heating of the spin system is of the order of 
\begin{align}
\tau^\text{spin}_h \sim  \frac{U}{\omega^2}.\label{eq:tau_h_spin}
\end{align}

\subsection{Heating of the charge sector}
\label{sec:heatingCharge}
Does the rising spin-temperature pose a problem for the operation of the Thouless pump?  How long does it take to create charge excitations, which are detrimental for the quantization of charge transport?

Here, we have to take two mechanisms into account: (i) heating of the spin sector and transfer of energy from the spin to the charge degrees of freedom and (ii) direct heating of the charge sector.

We first focus on the first mechanism, which dominates for small $\omega$. 
As long as the temperature is smaller than the charge gap, $T(t)\ll \Delta_c$, heating and effects arising from the thermalization of spin and charge are exponentially suppressed.

The spin-charge coupling becomes particularly relevant once the spin temperature approaches the charge gap scale, $T(t)\sim \Delta_c$. While spin and charge dynamics decouple at $U=\infty$ and for idealized Luttinger liquids, in the real system the spin- and charge sectors couple by terms arising both from interactions and also from the band curvature \cite{Haldane_1981}.
We can expect, for example, that there is a coupling of the type
\begin{align} 
H_{sc} \sim  \sum_j \tilde g_{sc}(t)\Psi_2^\dagger\Psi_1 \,\vec S_j \cdot \vec S_{j+1}+ h.c.
\end{align}
where $\Psi_{1/2}^\dagger$ creates charge excitations in the lower/upper instantaneous band of the Thouless pump. Such terms allow spin excitations to create particle-hole excitations across the charge gap.

These coupling terms are, however, strongly constrained by energy conservation.
A heating process requires that a large energy of order $\Delta_c \gg \bar J_1$ is transferred from the spin to the charge degrees of freedom. Due to energy conservation, this is only possible in a high-order  many-particle process of order $N_{cs}$ in the spin-charge coupling.

Exactly this problem has been studied extensively in the context of the lifetime of doublon excitations in Mott insulators \cite{doublonsLifetimeExp}. Based on these results, we estimate
\begin{align}N_{cs} \sim \frac{\Delta_c}{\bar J_1}.\end{align}
Thus, the rate of creating charge excitations for $T(t)\gtrsim  \Delta_c$
is exponentially small in $N_{cs}$,
\begin{align}\label{eq:Gamma_cs}
\Gamma_{cs} \sim e^{-\alpha' N_{cs}},
\end{align}
where the dimensionless prefactor $\alpha'$ depends logarithmically on system parameters as has, e.g., been discussed in Ref.~\cite{doublonsLifetimeExp}. 

We conclude, that in the infinite system charge transport remains quantized up to exponentially small corrections  for a long time after switching on the pump. Heating effects of the charge sector can be neglected for 
\begin{align}
t \ll \max\left(\frac{1}{\Gamma_{cs}},\tau^\text{spin}_h \right).
\end{align}
The maximum appears because spin-mediated charge heating requires both a sufficiently hot spin sector and sufficient time for spin-charge energy transfer. 

Furthermore, there are also direct heating processes of the charge sector arising from the simultaneous absorption of $N_{cc}$ Floquet energy quanta $\hbar \omega$ from the time-dependent Hamiltonian with $N_{cc} \hbar\omega \gtrsim \Delta_c$ or
\[ N_{cc}\sim \frac{\Delta_c}{\omega}\]
in units where $\hbar=1$.
Again, one obtains an exponential suppression of the direct charge-heating rate,
\begin{align}
\Gamma_{cc} \sim e^{-\alpha'' N_{cc}}.
\end{align}
Taking both mechanisms into account, the time scale for the heating of the charge sector is set by 
\begin{align}
\tau^\text{charge}_h \approx \min\!\left[
\frac{1}{\Gamma_{cc}},\max\!\left(\frac{1}{\Gamma_{cs}},\tau_h^\text{spin}\right)
\right].
\end{align}

For $\omega\to 0$ at finite $U$  the spin-heating time $\tau_h^{\rm spin}$  nominally dominates, \begin{align}
\tau^\text{charge}_h \sim \tau^\text{spin}_h \sim \frac{U}{\omega^2},
\end{align}
where we assumed as above the $\delta t\sim \delta v \sim t_0$.
But as the coupling of spin and charge $\Gamma_{cs}$ and the direct charge heating rate $\Gamma_{cc}$ are exponentially small for large $U$, they are expected to be  often the limiting factor deep in the Mott insulating phase.

For $\omega<\bar J_1$, the spin-charge channel gives the smaller exponent, $N_{cs}<N_{cc}$, and therefore limits the lifetime. In this case, we obtain for large $U$
\begin{align}
\tau^\text{charge}_h \sim e^{c \frac{U}{t_0}},\label{eq:tau_charge}
\end{align}
where $c$ depends logarithmically on system parameters.

The numerical results in Fig.~\ref{fig:pump_vs_U} reveal an additional effect that is not captured by the preceding discussion which focusses on the thermodynamic limit. According to our numerics, the deviations from quantized charge pumping are largest for low-energy states and smallest for high-energy states. This observation is counterintuitive: in the data shown here, a spin sector that is initially ``colder'' is more efficient at generating charge-sector excitations than spin sectors that are initially more highly excited. In a loose sense, the effect is reminiscent of the (quantum) Mpemba effect~\cite{Ares2025}, where a state that is initially farther from equilibrium may relax faster than one that is initially closer to equilibrium. While we do not fully understand the origin of this finite size effect, we note that the highest energy state of an antiferromagnetic Heisenberg model is the fully spin-polarized state. For a fully polarized state there is no inelastic scattering and no heating, no conversion from spin to charge and thus deviations from the quantized value are the smallest. Thus, deviations from quantization get smaller upon increasing the energy. We expect that this physics is not important in the thermodynamic limit, where states with large spin polarization are exponentially rare, see App.~\ref{app:numerics} for further discussions.


\begin{figure}
    \centering
    \includegraphics[width=0.6\linewidth]{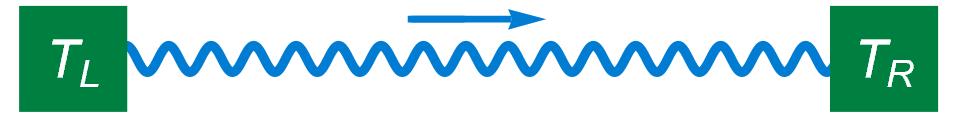}
    \caption{Schematic picture of a Thouless pump operating between two reservoirs.}
    \label{fig:pump_schematic}
\end{figure}

\subsection{Heating of a finite-size Thouless pump}

Above, we considered an infinite system but for the operation of an actual pump, one should instead consider a finite system of length $L$ operating between two reservoirs, see Fig.~\ref{fig:pump_schematic}.
Assuming that the reservoirs are at fixed temperature, $T_L=T_R$, we can expect that the system reaches a steady state. Here,  cooling arises due to the coupling to the reservoirs.

For large $L$ and small $\omega$, heat in the spin system is described by the drift-diffusion equation
\begin{align}
\partial_t e+v \partial_x e-D \partial_x^2 e=- \gamma_h e,
\label{eq:drift_diffusion}
\end{align}
where $D$ is the energy diffusion constant. The drift term with  $v=a \frac{\omega}{2 \pi}$ arises because the operation of the pump transports the spin background on average by one lattice constant per cycle. $\gamma_h \sim \omega^2 \frac{(\delta J)^2}{\bar J_1^3}$ parametrizes the bulk heating rate according to Eq.~\eqref{eq:heatingE}. Note that  $e \sim -\frac{\bar J_1^2}{T}$, Eq.~\eqref{eq:T},
and heating corresponds to $\gamma_h>0$, see Eq.~\eqref{eq:heatingE}.

\begin{figure}
    \centering
    \includegraphics[width=0.6\linewidth]{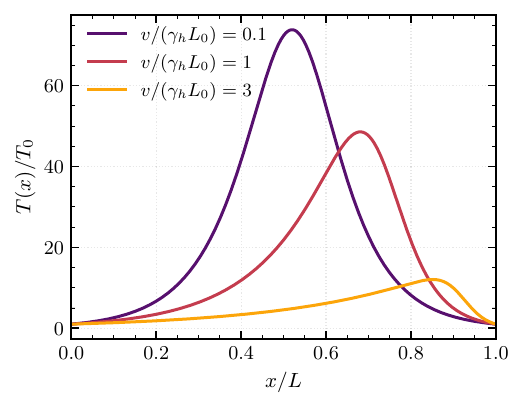}
    \caption{Steady-state temperature profile obtained from the solution of the drift-diffusion equation (Eq.~\ref{eq:drift_diffusion}) for different values of the dimensionless parameter $v/(\gamma_h L_0)$. The asymmetric profile arises as our system effectively pumps charge, spin and heat from left to right with the velocity $v$. Here, we use $L=10L_0$.}
    \label{fig:T_profile}
\end{figure}

The drift-diffusion equation can be solved analytically for a finite system of length $L$ with the boundary condition $e(0)=e(L)=\epsilon_0<0$, or, equivalently $T_L=T_R=T_0$. 
For the highest spin-temperature along the wire, $T_{\text{max}}$, we obtain for $t\to \infty$ an exponential growth for large $L$
\begin{align}
\frac{T_{\text{max}}}{T_0} \approx \left\{\begin{array}{ll}
    \cosh\!\left(\frac{L}{L_0}\right) & \text{for } L_0 \gg v/\gamma_h \\
    e^{L \gamma_h/v}  & \text{for } L_0 \ll v/\gamma_h 
\end{array}\right. .
\end{align}
where $L_0 \equiv \sqrt{D/\gamma_h}$.
In Fig.~\ref{fig:T_profile}, we plot the solution for the steady-state temperature profile showing the evolution as a function of $v/(\gamma_h L_0)$.
As $\gamma_h \sim \omega^2$ and $v\sim \omega$,
 the exponential growth is in both limits linear in $\omega$,
\begin{align}
\frac{T_{\text{max}}}{T_0} \sim e^{\frac{L \omega}{v_0}}
\end{align}
with $v_0\sim \max\!\left(a  \frac{\bar J_1 ^3}{ (\delta J)^2}, \left(\frac{\bar J_1^3 D}{(\delta J)^2}\right)^{1/2} \right)$.

Note that the energy diffusion constant $D$ is {\em not} of order $ a^2 \bar J_1 $ as suggested by dimensional analysis but much larger.
In the integrable nearest-neighbor Heisenberg model, the heat current is exactly conserved~\cite{zotos1997}, implying ballistic energy transport and formally infinite thermal diffusion constant $D$. Thus, integrability breaking terms determine the size of $D$.

It was shown in Ref.~\cite{Jung2006} both numerically and analytically that the leading-order corrections to the decay-rate of the heat current, $\propto (\bar J_2)^2$, vanish exactly in the static limit resulting in $D\sim 1/(\bar J_2)^4$. Such a cancellation of the second-order term is absent for finite $\omega$ \cite{Jung2006}. We thus conclude that
\begin{align} 
D \sim a^2 \frac{(\bar J_1)^5}{\max\!\left[(\bar J_2)^4,(\delta J)^2 \omega^2\right]}.
\end{align}
Thus, $D\gg \frac{v^2}{\gamma_h} \sim \frac{(\bar J_1)^3}{(\delta J)^2}$
resulting in 
\begin{align}\label{tmax}
\frac{T_{\text{max}}}{T_0} \sim e^{L \sqrt{\gamma_h/D}} \sim e^{ \alpha' L \omega \frac{|\delta J| \text{max}\left[(\bar J_2)^2,|\delta J|\omega\right]}{a (\bar J_1)^4} }\sim e^{L/\ell_h^\text{\,spin}},
\end{align}
where $\alpha'$ is a number of order 1 and we introduced the length scale $\ell_h^\text{\,spin}$ which controls the heating of the spin system.

As both $\delta J$ and $\bar J_2$ scale with $1/U^2$ while $\bar J_1 \sim 1/U$, we therefore
estimate for $\delta t\sim \delta v\sim t_0$
\begin{align}\label{Llower}
\ell_h^\text{\,spin} \sim a\, \text{min}\!\left[\frac{U^2}{\omega t_0},\frac{t_0^2}{\omega^2}\right].
\end{align}

Next, we have to investigate whether spin and charge equilibrate. The particles move from the left to the right reservoir during the time
   $ \Delta t = \frac{L}{a} \frac{2 \pi}{\omega}$. Spin and charge equilibrate for 
$\Gamma_{cs} \Delta t \ge 1$, where $\Gamma_{cs}$ is the charge-spin relaxation rate introduced in Eq.~\eqref{eq:Gamma_cs}. Thus, the characteristic length scale for the equilibration of charge and spin is of the order of 
\begin{align}
\ell_{cs} \sim a \frac{\omega}{\Gamma_{cs}}.
\end{align}
Finally, the quantized pumping of our incoherent Thouless pump is  affected by heating only if two effects occur simultaneously: (i) the spin temperature gets of the order of the charge gap and (ii) the spin- and charge sectors start to equilibrate.
Therefore, the length scale for heating of the charge sector is set by the larger of the two length scales for spin-heating and spin-charge equilibration,
\begin{align}\label{eq:l_charge}
\ell_h^\text{\,charge}&\sim \max(\ell_h^\text{\,spin},\ell_{cs})\\
&\sim a\,\text{max}\!\left( \text{min}\!\left[\frac{U^2}{\omega t_0},\frac{t_0^2}{\omega^2}\right],\frac{\omega}{\Gamma_{cs}}\right),\nonumber
\end{align}
where we ignore logarithmic corrections to the prefactor obtained by solving, e.g., the equation $T_\text{max}\sim \Delta_c$, Eq.~\eqref{tmax}, for $L$.
For $L\ll \ell_h^\text{\,charge}$, we predict that the  incoherent Thouless pump operates with exponential precision. For $L\gtrsim \ell_h^\text{\,charge}$, in contrast, the deviations from quantized pumping become substantial  due to the presence of unwanted charge excitations.

In the adiabatic limit, $\omega \to 0$, 
the bottleneck is given by the first term in Eq.~\eqref{eq:l_charge} with $\ell_h^\text{\,charge} = \ell_h^\text{\,spin}\sim a \frac{U^2}{\omega t_0}$.
For finite $\omega$ and large $U$ instead, there is a large regime of parameters where the incoherent Mott-Thouless pump is  protected by the exponential suppression of spin-charge coupling, $\ell_h^\text{\,charge}\sim e^{c U/t_0} $.

In conclusion, incoherent Mott-Thouless pumps can operate efficiently  over parametrically large system sizes.

\section{Heating in the continuum model}\label{sec:heating_continuum}
All estimates in the above sections have been obtained for the model Hamiltonian Eq.~\eqref{eq:H1} in the limit of large $U$. However, very similar results can be obtained for the continuum model, Eq.~\eqref{eq:H2} for finite but small $V_0-V_1$. After a Galilean transformation to the comoving frame, one obtains a time dependent potential
\begin{align}
\delta V(x,t)=\delta V \cos(k_0 x + 2 \omega t)
\end{align}
with $\delta V=\frac{V_0-V_1}{2}\ll V_0$ assumed to be much smaller than the amplitude of the potential $V_0 \cos(k_0 r)$ which is static in the comoving frame. The main effect of $\delta V \ll V_0$ is that it leads to a periodic oscillation of the height of the periodic potential, \begin{align}V_\text{max}\approx V_0+\delta V \cos(2  \omega t).\end{align} This naturally induces an oscillation in the tunneling amplitudes between neighboring minima. 

As discussed in Sec.~\ref{sec:models}, we assume that $V_0$ is large compared to the recoil energy, $E_R=\frac{(2 \hbar k_0)^2}{2 m}$.
In this limit, the tunneling rate can be computed from a WKB approximation.
Up to algebraic prefactors, the tunneling rate obtained from a WKB approximation scales as
\begin{align}
    t_\text{eff} &\sim e^{-2 \sqrt{V_\text{max}/E_R}} \sim e^{-2 \sqrt{V_0/E_R}} e^\Delta \nonumber \\
     \Delta&=\frac{\delta V \cos(2 \omega t)}{\sqrt{E_R V_0}}.
\end{align}
Thus, the ratio $\Delta$ controls the amplitude of the oscillations of the tunneling rate.
For $\Delta \ll 1$, the relative changes in the tunneling rate are small. In contrast for $\Delta \gg 1$, the relative changes are exponential in $\Delta$. 

Assuming that the tunneling rates are small compared to $U$, we can map the spin-sector of the model to a Heisenberg model with time-dependent nearest and next-nearest neighbor couplings $J_1$ and $J_2$ with $J_2\sim t_\text{eff}^4/U^3$.

For simplicity, we focus on the regime where the amplitude of $\Delta$ is of order one, $\Delta \sim 1$, which corresponds to the case $\delta t\sim \delta v\sim t_0$ which we used in the previous sections. In this case, the main qualitative differences between the two models arise because $J_2$ and the charge gap are of order $t_\text{eff}^4/U^3$ and $U$, respectively, in the continuum model, while they are of order $t_0^3/U^2$ and $t_0$ in the interacting Rice-Mele model.
 Similarly, the exponential suppression of the spin-charge coupling is of the order of 
$N_{cs}\sim \frac{U^2}{t_\text{eff}^2}$ instead of $N_{cs}\sim\frac{U}{t_0}$.

Thus, both the spin and charge sectors exhibit longer heating times:
\begin{align}
    \tau_h^\text{spin} \sim \frac{U^3}{\omega^2 t_\text{eff}^2}, \qquad \tau^\text{charge}_h \sim e^{c \frac{U^2}{t_\text{eff}^2}},
\end{align}
instead of the scalings in Eq.~\eqref{eq:tau_h_spin} and Eq.~\eqref{eq:tau_charge}, respectively.

\begin{figure}[t]
    \centering
    \includegraphics[width=0.6\linewidth]{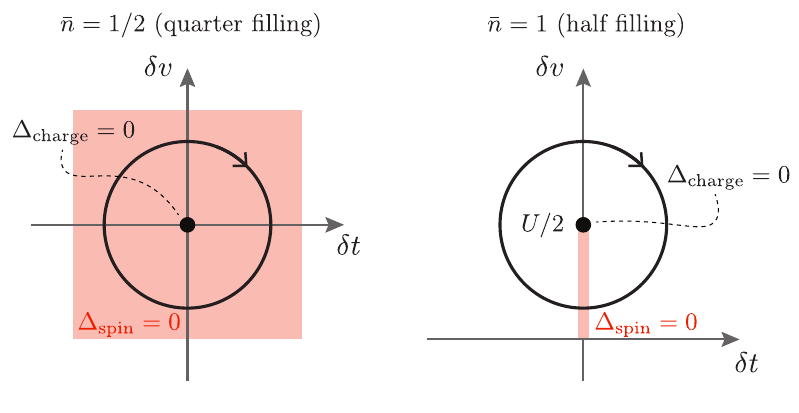}
    \caption{Operation of Mott-Thouless pumps. Comparison of the quarter filled and half-filled Ref.~\cite{Walter2023,Viebahn2024} case, driven by periodic changes of the staggered potential and the staggered hopping (circles). The cycle encloses a point (center) where the charge gap closes. The red shades shows regions where the spin-gap closes.}
    \label{fig:pump2}
\end{figure}

\section{Mott-Thouless pump at half filling}\label{sec:half}
Above, we considered a Mott-Thouless pump in a quarter-filled model. Recently, an experiment by the Esslinger group \cite{Walter2023,Viebahn2024} reported a new type of Thouless pump. While based on the same Hamiltonian, Eq.~\eqref{eq:H1}, it studied a half-filled Hubbard model instead.
While in the non-interacting Thouless pump two fermions are pumped by $2a$, 
the authors showed both numerically  and experimentally that for strong interactions and a certain pumping protocol one can obtain  a pump transporting a single charge by two lattice constants (similar to the $1/4$-filled Mott-Thouless pumps). Importantly, Ref.~\cite{Viebahn2024} observed quantized pumping only for a single pumping cycle and the authors showed its breakdown for the next cycle or for modified starting conditions.  Repeating some of the arguments of that paper, we would like to explore why the breakdown happens and why the system is different 
from the quarter filled case. 

\subsection{Charge and spin pumping at half filling}
\begin{figure}[th]
    \centering
    \scalebox{0.99}{%
        \begin{tabular}{@{}c@{\hspace{2pt}}c@{\hspace{12pt}}c@{\hspace{2pt}}c@{}}
            \raisebox{\dimexpr\ht\strutbox-\height}{\textbf{(a)}} &
            \raisebox{\dimexpr\ht\strutbox-\height}{\includegraphics[width=0.42\linewidth]{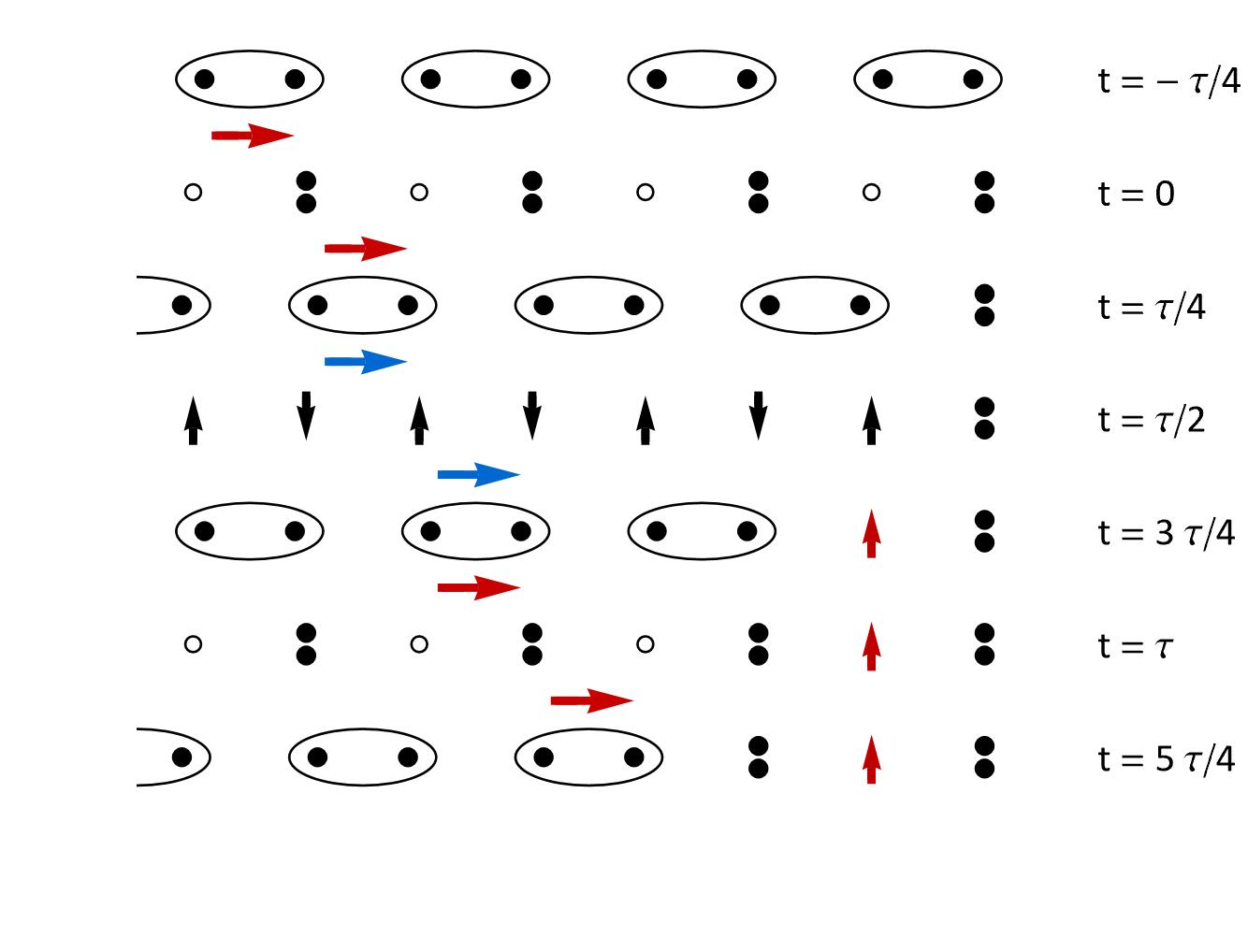}} &
            \raisebox{\dimexpr\ht\strutbox-\height}{\textbf{(b)}} &
            \raisebox{\dimexpr\ht\strutbox-\height}{\includegraphics[width=0.42\linewidth]{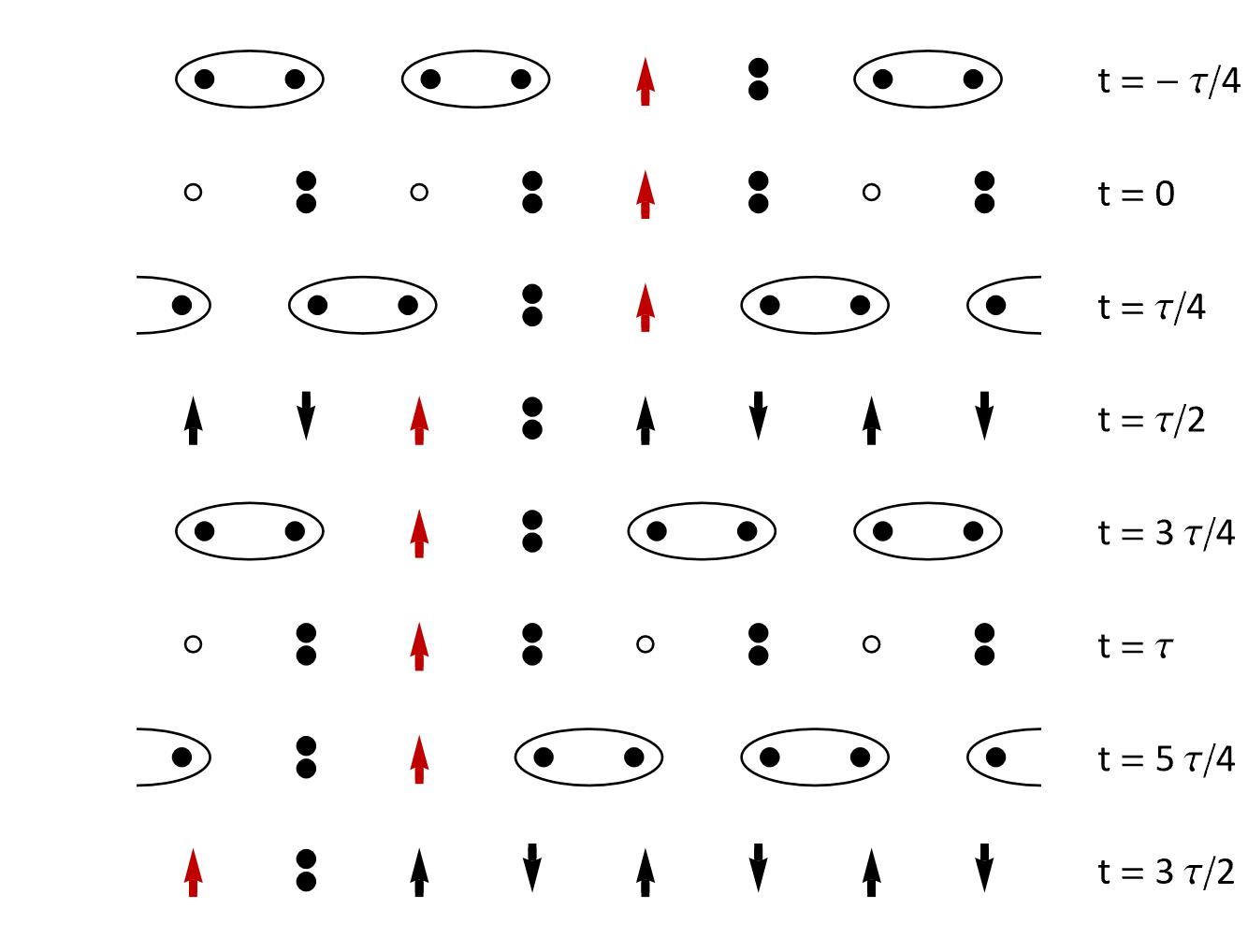}} \\[3ex]
            \raisebox{\dimexpr\ht\strutbox-\height}{\textbf{(c)}} &
            \raisebox{\dimexpr\ht\strutbox-\height}{\includegraphics[width=0.42\linewidth]{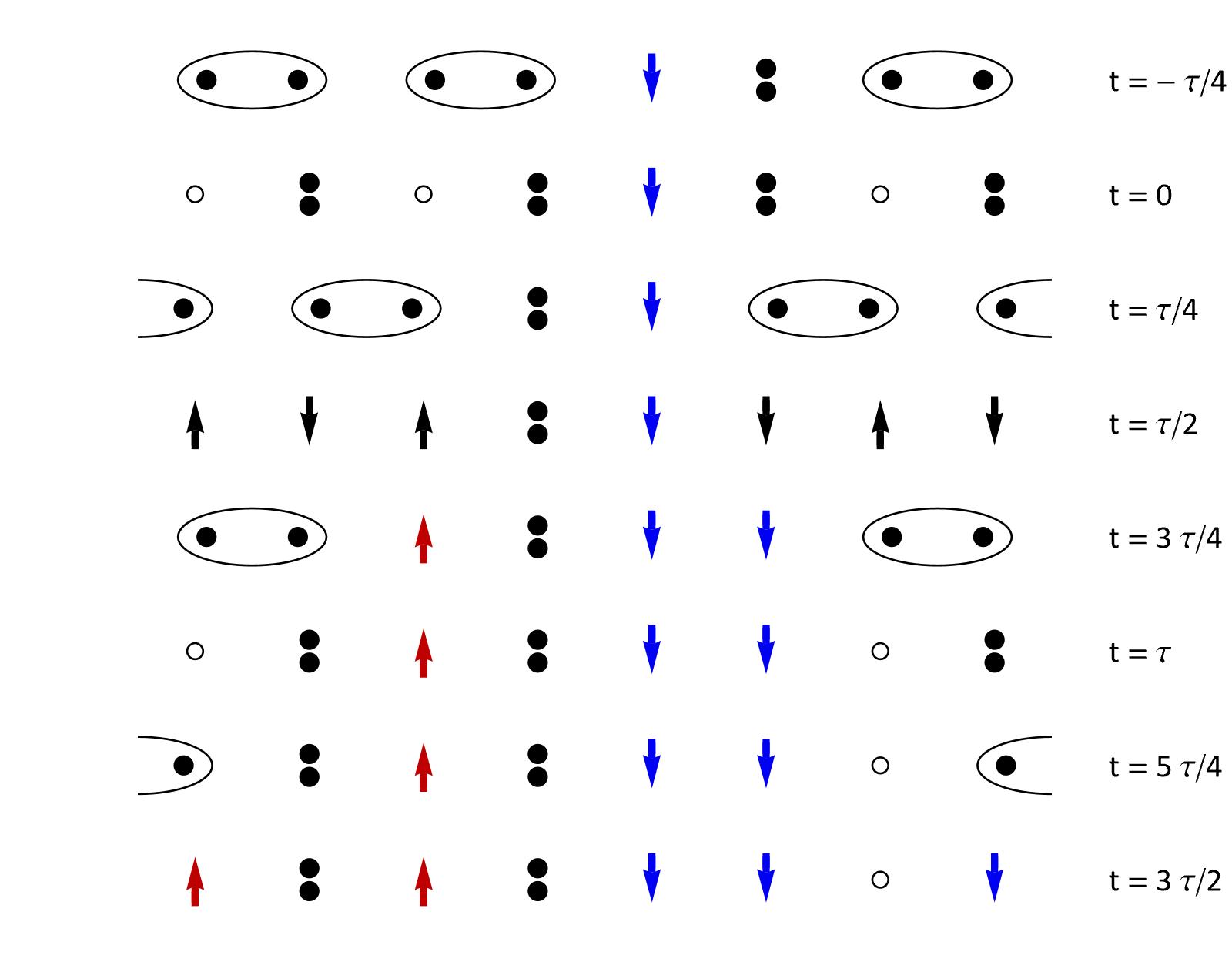}} &
            \raisebox{\dimexpr\ht\strutbox-\height}{\textbf{(d)}} &
            \raisebox{\dimexpr\ht\strutbox-\height}{\includegraphics[width=0.42\linewidth]{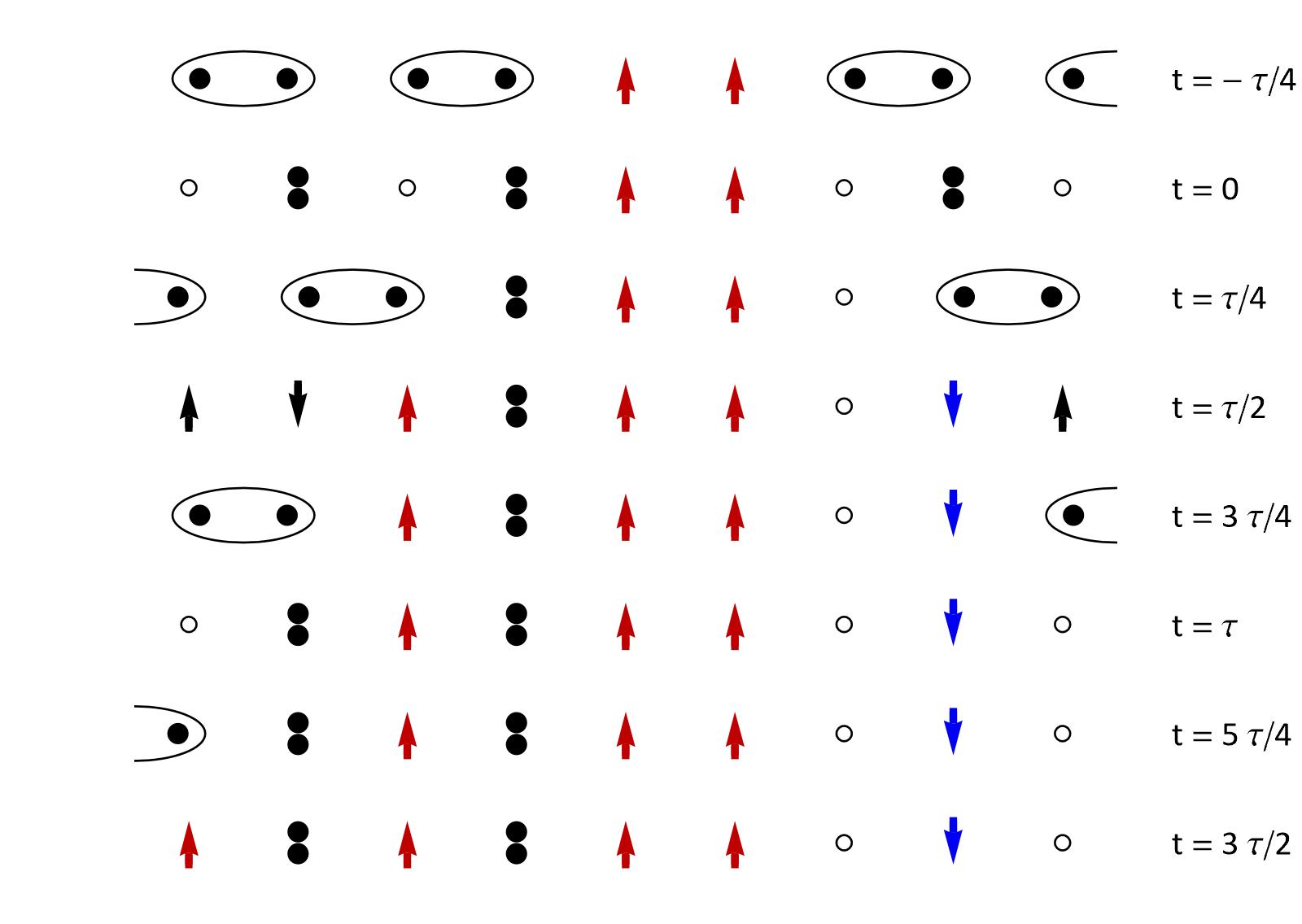}}
        \end{tabular}%
    }
    \caption{Operation and defect propagation in the Mott-Thouless pump at half filling. 
    (a) Result of a calculation within the simplified model $H_s$, where hopping acts only on even (odd) sites for $0<t<\tau/2$ ($\tau/2<t<\tau$) for a semi-infinite chain with open boundary conditions on the right. Filled dots, circles and arrows denote charges, singlet states and localized spins, respectively. The red and blue horizontal arrow denotes the transport of charge and spin, respectively. For the chosen staggered magnetic field the transported fermion carries spin-up only. Charge and spin accumulate at the right boundary. 
    (b) Propagation of low-energy defects within model $H_s$: An added spin-$\uparrow$ fermion moves to the left, opposite to the motion of the Thouless pump. 
    (c) An added spin-$\downarrow$ fermion is localized but cuts the chain into two pieces. Therefore, charge and spin-$\uparrow$ accumulate to the left while holes and spin-$\downarrow$ accumulates to the right. 
    (d) A spin excitation also blocks the Thouless pump, inducing a similar accumulation of spin and charge as in panel (c).} 
    \label{fig:pumpD}
    
    \makeatletter
    \protected@edef\@currentlabel{\thefigure a}\label{fig:pump_a}
    \protected@edef\@currentlabel{\thefigure b}\label{fig:pump_b}
    \protected@edef\@currentlabel{\thefigure c}\label{fig:pump_c}
    \protected@edef\@currentlabel{\thefigure d}\label{fig:pump_d}
    \makeatother
\end{figure}

The operation of the pump introduced in Ref.~\cite{Walter2023,Viebahn2024} is best understood in the strong coupling limit of our model, Eq.~\eqref{eq:H1}.
For vanishing or small hopping, $t_0,\delta t \ll U,\delta v$,
one obtains a Mott insulating phase when the staggered potential is small compared to $U/2$, $\delta v < U/2$. This phase has a charge gap but no spin gap. In contrast, for  $\delta v > U/2$, one obtains a fully gapped charge-density wave as sketched in the second row of Fig.~\ref{fig:pump_a}. To realize the pump, one moves in parameter space around the quantum phase transition between those two phases at finite $t_0 \sim \delta t\ll U$ using, for example,
\begin{align}
\delta t(t)=\delta t \sin(\omega t),\qquad \delta v(t) = \frac{U}{2}+\delta v \cos(\omega t)
\label{eq:drive_half}
\end{align}
for the staggered hopping and staggered potential.
The main differences to Eq.~\eqref{path1}
are the large, constant offset $U/2$ in $\delta v(t)$ and, importantly, that the system is operated at half filling instead of quarter filling.
Furthermore, we will also discuss the role of a small staggered magnetic field,
\begin{align}
\Delta H(t)= \delta B(t) \sum_{\alpha,\beta=\uparrow/\downarrow,i} (-1)^i c^\dagger_{i,\alpha} \sigma^z_{\alpha,\beta} c_{i,\beta}.
\end{align}
Such a staggered field opens a  gap at $t=\tau/2$, when
$\delta t=0$, $\delta v < U/2$, when the system maps to a Heisenberg chain. In the presence of this term, one has a fully gap-protected Thouless pump while a gapless state is unavoidable for $\delta B=0$. While such a staggered field  was not considered in the experimental system and the theoretical analysis of Ref.~\cite{Viebahn2024}, we will show that its presence allows for an easier interpretation of the physics underlying the topological pump.

To obtain an analytic understanding on how the pump operates, it is useful to {\em adiabatically deform} the pumping cycle $H_\text{RM}(t)$ to a simplified  model $H_s$, where the staggered hopping $t_0+(-1)^i \delta t \sin(\omega t)$
is replaced by a hopping term which for $0\le t \le \tau/2$ ($\tau/2 < t \le \tau$) is only active for even (odd) sites, respectively. 
\begin{align}H_s:\quad
\begin{array}{llll}
t^o&=0, \qquad &t^{e}=(t_0+\delta t) \sin(\omega t) \quad & \text{for } 0\le t <\tau/2 \nonumber \\
t^e&=0, \quad &t^o=- (t_0+\delta t) \sin(\omega t) \quad & \text{for } \tau/2\le t <\tau.
\end{array}
\end{align}
Here, the hopping from site $i$ to site $i+1$ is given by $t_i=\frac{1}{2} \left(t^e+t^o+(-1)^i (t^e-t^o)\right)$.
To obtain a unique ground state at all times, we use  a small staggered field $\delta B$
which, for simplicity, we assume to be finite (and small) only in a small interval around $t= \tau/2+n \tau$, see below.

It is straightforward to solve $H_s(t)$ exactly in the adiabatic limit, $\tau=\frac{2 \pi}{\omega} \to \infty$. The resulting solution is sketched for the ground state in Fig.~\ref{fig:pump_a}. For $t=\pm \tau/4+n \tau$, $n \in \mathbb Z$, the ground state is a product states of singlets on even or odd links, respectively. At $t=0+n \tau$, where $\delta v > U/2$, one obtains a charge-density wave. At $t=\tau/2+n \tau$, in contrast,  $\delta v < U/2$ with $t^e=t^o=0$, one is in a Mott-insulating phase, where the staggered magnetic field induces the antiferromagnetic state sketched in the fourth row of Fig.~\ref{fig:pump_a}.

One can now track the transport of charge and spin for one pumping cycle by a visual inspection of Fig.~\ref{fig:pump_a}. Between the first and second row, $-\tau/4<t<0$, but also between the second and third row, $0<t<\tau/4$, one unit of charge is transported to the right by a distance of $a_0$ (red arrow). For $\tau/4<t<\tau/2$ and also for $\tau/2<t<\tau$, there is no charge transport but a net transport of spin-up (blue arrow). The net effect is that per pumping cycle one spin-up particle is transported by one unit-cell size, $a=2 a_0$, while there is no transport of spin-down particles for the chosen sign of the product $\delta v \delta B$. As we consider in Fig.~\ref{fig:pump_a}
an open boundary condition on the right side, pumping leads to an accumulation of spin-up particles at the right boundary.

The net transport of one charge per pumping cycle fits to the value reported in Ref.~\cite{Viebahn2024} but due to the absence of  $\delta B$, spin-up and spin-down were each transported with $50\%$ probability.

\subsection{Pumping in the presence of spin and charge defects}\label{subsec:defects}

For $\delta B=0$ the pumping cycle necessarily crosses a point inside the Mott-insulating phase where the spin-gap closes while the charge gap remains finite. Thus, the pumping cycle will necessarily create spin excitations per cycle (at $T=0$ the corresponding rate can be computed from a Kibble-Zurek formula \cite{Grandi2010}).

We will show below, that these spin excitations are {\bf not} harmless in the half-filled case as they necessarily create charge excitations. This is opposite to our findings for the quarter-filled case reported in Sec.~\ref{heatingQuarter}.

In Fig.~\ref{fig:pumpD}, we show how three types of defects propagate within the simplified model system $H_s$. For each type of defect we study eigenstates of $H_s$ with the lowest energy during an adiabatic pumping cycle.

An added spin-up particle (Fig.~\ref{fig:pump_b}) moves per pumping cycle by $a=2 a_0$ to the left, in a direction opposite to the pumping direction of the Mott-Thouless pump. Removing a spin-up particle (not shown) has a similar effect: it creates a hole which moves per pumping cycle by $a=2 a_0$ to the right. In both cases, charge moves in the `wrong' direction, thus destroying the quantization of the half-filled Mott-Thouless pump.

An added spin-down particle (Fig.~\ref{fig:pump_c}) has a different effect. It effectively cuts the chain and acts as a barrier mainly because at $t=\tau$ the Pauli principle prohibits the formation of a doubly-occupied site as has also been argued in Ref.~\cite{Viebahn2024}. As the Mott-Thouless pump continues to operate to the left and to the right of this defect, per pumping cycle an extra spin-up accumulates to the left of the defect 
while a spin-up electron is removed from the right-side of the defect. The same happens (not shown) when a spin-down particle is removed rather than added to the system.

Finally, we consider the role of a spin-flip excitation, replacing in the first row of Fig.~\ref{fig:pump_d} a singlet by a spin-up triplet. The defect is localized and acts as a barrier creating in each cycle charges on the left and holes on the right as has also been pointed out in Ref.~\cite{Viebahn2024}. Exactly the same result is obtained when the initial $\uparrow\!\uparrow$ is replaced by a $\downarrow\!\downarrow$ state, as can also be seen from Fig.~\ref{fig:pump_c} where such a state emerges for $t\ge 3 \tau/4$.

\subsection{Conversion of spin into charge excitations and comparison to the experiment}
The results of Sec.~\ref{subsec:defects} and Fig.~\ref{fig:pumpD} apply directly only to the simplified model $H_s$ and some of the results are  artifacts of this exactly solvable model. For example, the localization of defects in space will not occur in a model where both even and odd hoppings are finite as in $H_\text{RM}$, Eq.~\eqref{eq:H1}.

Nevertheless, one central observation carries over to the full model: during the adiabatic cycle it is possible that a pure spin excitation is converted into a pair of charge excitations, a hole and a doubly occupied site. As this is possible within $H_s$, it will also occur (perhaps with a modified rate) in the more generic model $H_\text{RM}$ at half filling. At the same time, during a pumping cycle the spin gap necessarily closes (in the absence of a staggered magnetic field) and there is no exponential protection against generating spin-excitations.
We therefore conclude that at half filling there is no exponential protection of topological charge pumping in the half-filled model and the rate of creating charge defects from spin defects is of order $1$ per cycle, \begin{align}
    \Gamma_{cs}\sim \omega.
\end{align}

The coupling between spin and charge in the half-filled model can also be seen directly using Abelian bosonization. We decompose the low-energy lattice fermion into right- and left-moving fields,
\[
\Psi_\sigma(x) \simeq e^{i k_F x}\psi_{R,\sigma}(x)
+e^{-i k_F x}\psi_{L,\sigma}(x),
\]
which are bosonized using the convention~\cite{Giamarchi2004}
\begin{align}
\psi_{R/L,\sigma}(x)
\sim
e^{
-\frac{i}{\sqrt2}
\left[
\pm\phi_c(x)-\theta_c(x)
+\sigma\left(\pm\phi_s(x)-\theta_s(x)\right)
\right]
}.
\end{align}
At half filling, the staggered potential and staggered hopping generate a single-particle backscattering term,
\begin{align}
    H_B \propto \int dx\,
    \left(v_0+v_1(t)e^{i\alpha(t)}\right)
    \sum_\sigma \psi_{R\sigma}^\dagger \psi_{L\sigma}
    + h.c.,
\end{align}
where $v_0$ denotes the time-independent component of the staggered potential and $\alpha(t)=\omega t$ is the relative phase of staggered potential and staggered hopping. Bosonizing this term gives
\begin{align}
    H_B \sim \!\!\int \! dx\,
    \left[
    v_0 \cos(\sqrt{2}\phi_c)
    +v_1 \cos(\sqrt{2}\phi_c-\alpha)
    \right]\!
    \cos(\sqrt{2}\phi_s).
\end{align}
Thus the perturbation that gaps the charge sector also depends explicitly on the spin field. A local spin excitation, corresponding to a kink that shifts $\phi_s$ by $\pi/\sqrt{2}$, changes the sign of $\cos(\sqrt{2}\phi_s)$ and therefore reverses the effective potential acting on the charge field $\phi_c$, thus creating  charge excitations. This illustrates the strong spin-charge coupling in the half-filled pump.

The situation is different in the quarter-filled model where the corresponding single-particle backscattering term is forbidden by momentum conservation. Instead, the combination of interactions with the staggered potential and/or staggered hopping generates a time-dependent Umklapp term,
\begin{align}
    H_U \propto \int dx\,
    e^{i\alpha(t)}
    \psi_{R\uparrow}^\dagger \psi_{R\downarrow}^\dagger
    \psi_{L\downarrow}\psi_{L\uparrow}
    + h.c.,
\end{align}
which bosonizes \cite{Giamarchi2004,Rosch2000} to
\begin{align}
    H_U \propto \int dx\, \cos(2 \sqrt{2} \phi_c-\alpha(t)).
\end{align}
This term lacks the coupling of spin- and charge of the quarter-filled model.

The conversion of spin into charge excitations  in the half-filled model is also consistent with the experimental and numerical results reported by Viebahn {\it et al} in Ref.~\cite{Viebahn2024}, see  App.~\ref{app:numerics} for further numerical data. While in the first pumping cycle an approximately quantized transport of charge is reported, pumping breaks down almost completely in the second pumping cycle, see Fig.~8b of Ref.~\cite{Viebahn2024} and App.~\ref{app:numerics}. Similarly, also the numerical results reported in Ref.~\cite{Viebahn2024} show a complete breakdown of quantized pumping after the first pumping cycle, Fig.~13 in Ref.~\cite{Viebahn2024}. Remarkably, the authors could show that multi-cycle quantized pumping is restored as soon as one opens a large spin gap by adding a large nearest-neighbor Ising interaction to the model which completely suppresses the formation of spin excitations and acts similar to our staggered magnetic field.  

Thus, a consistent picture emerges: at half-filling spin excitations efficiently destroy quantized charge pumping in strong contrast to our results for quarter filling.

\section{Conclusion}
Many theoretical approaches in condensed matter physics are formulated in terms of coherent quantum states, quasiparticles, and low-entropy ground states. In contrast, highly entropic incoherent regimes remain substantially less explored, both conceptually and theoretically, as they often evade descriptions based on mean-field approximations, quasiparticle pictures, or adiabatic continuity from the ground state.

Nevertheless, highly incoherent many-body systems can exhibit novel forms of universal collective behavior, as discussed, for example, in the context of pseudogap physics in the cuprates~\cite{Haule2002,HauleEDMFT}, spin-incoherent one-dimensional systems~\cite{Fiete07,FeiguinFiete2010}, or in studies of incoherent metals~\cite{Hartnoll2015} and SYK-type models characterized by the absence of quasiparticles and by extensive residual entropies \cite{SYKreview}.

In this work, we showed that topologically quantized transport can survive in highly incoherent regimes with large thermodynamic entropies. We demonstrate that quantized Thouless pumping remains exponentially protected in spin-incoherent Mott states despite their extensive entropy. The key mechanism is a dynamical separation of sectors: while the spin degrees of freedom are incoherent and carry a large entropy, the transported charge resides in a gapped sector that remains effectively decoupled from the spin dynamics.

Our results are particularly relevant for ultracold-atom experiments, where the preparation of low-entropy fermionic many-body states remains a major challenge despite remarkable recent progress~\cite{Xu2025}. From this perspective, the existence of exponentially protected topological charge pumping operating deep in the spin-incoherent regime enlarges the parameter regime where topological phenomena can be observed experimentally.

At the same time, our analysis of the recently realized half-filled interacting Thouless pump~\cite{Walter2023,Viebahn2024} demonstrates that not all incoherent regimes are equally benign. In that system, spin excitations are converted into charge excitations, thereby destroying the exponential protection of quantized transport.

More generally, our work raises the question to what extent topological phenomena beyond Thouless pumping can survive deep in incoherent many-body regimes. In particular, it would be interesting to explore 
 further whether interacting Chern-insulating phases or other
topological states can retain robust quantized responses in high-entropy states. 
For example, it has been shown that a partially filled band of a Thouless pump \cite{Lindner2017,gulden2020} or a partially filled Chern bands \cite{aidelsburger2015measuring} exhibit quantized transport,
provided that all momentum states are equally populated -- a long-lived state reached once the particles have thermalized within the band.   

Our results suggest that robust topological phenomena do not fundamentally require low-entropy ground states, but may persist in highly entropic quantum matter whenever coherent transport sectors remain dynamically protected from incoherent degrees of freedom.

\section*{Acknowledgments}
We acknowledge useful discussions with Sebastian Diehl and, especially, with Fabian Heidrich-Meisner.
This work was funded by the Deutsche Forschungsgemeinschaft (DFG, German Research Foundation) through CRC/TR 183 (Project No.~277101999, subproject A1), CRC 1238 (Project No.~277146847, subproject C2 and C3), and the Emmy Noether program (U.F.P.S., Project No.~544397233). ChatGPT 5.5 (OpenAI) was used for language editing.
Numerical simulations were carried out using Python scripts developed with the assistance of Google Gemini.

\paragraph*{Code and Data availability}
The supporting data and codes for this article are available from Zenodo at \doi{10.5281/zenodo.22100125}~\cite{kumar2026zenodo}.

\bibliography{incoherentThouless.bib}%

\newpage
\begin{appendix}
\section{Effective Heisenberg model for spin degrees of freedom}\label{app:effective_Heisenberg}
This section discusses the effective Heisenberg model describing the spin excitations of the quarter-filled interacting Rice-Mele model at large $U$.

The  degenerate instanenous ground states of the model of Eq.~\eqref{eq:H1} at $U=\infty$ can be written in a first-quantized language as
\begin{align}
|\Psi_\chi(t)\rangle
=&
\sum_{x_1<\cdots<x_N}
\phi_{SL}^t(x_1,\dots,x_N)  \\
& \quad \times
\sum_{\sigma_1,\dots,\sigma_N}
\chi(\sigma_1,\dots,\sigma_N)\,
c^\dagger_{x_1\sigma_1}\cdots c^\dagger_{x_N\sigma_N}\,|0\rangle.\nonumber
\end{align}
where $\phi_{SL}$ is the Slater determinant describing the lowest filled band of the spinless non-interacting system and $H_\text{RM}|\Psi_\chi(t)\rangle=E(t) |\Psi_\chi(t)\rangle $ at $U=\infty$.  Here, $x_i$ are lattice coordinates and we assume open boundary conditions for simplicity. The spins are defined by an ordering in real-space, implying that spin $i$ is always left of spin $i+1$. The antisymmetry of the Slater determinant ensures that there are no double occupancies.
As per cycle of the Thouless pump, all charges move one unit cell, this also applies to the spin degrees of freedom. 

For finite $U$, on the other hand, virtual charge fluctuations lift the remaining degeneracy and endow the spins with their own dynamics governed by non-zero spin exchange couplings. 
The goal of this section is to obtain a spin-model description of the insulating state of the Rice-Mele model described in the main text via a Schrieffer-Wolff (SW) transformation. For convenience, we define the staggered hopping strengths $t_w \equiv t_0 - \delta t(t)$ and $t_s \equiv t_0 + \delta t(t)$, and assume $t_w < t_s$.
We will work with the following hierarchy of energy scales: $t_w < \Delta_W \ll U$, where $U$ is the on-site Hubbard interaction and $\Delta_W$ is the charge gap. 

Let us define the subspace without any doublon excitations as $\mathcal{V}_U$ and within $\mathcal{V}_U$, a subspace without charge-excitations denoted by $\mathcal{V}_{\Delta}$. 
The strategy is to first obtain an effective Hamiltonian within $\mathcal{V}_U$ to order $t_w^2/U$. As a second step, we will then derive an effective spin Hamiltonian $H_{\text{low}}$ within the low-energy subspace $\mathcal{V}_{\Delta}$, which is devoid of charge excitations. The effective Hamiltonian within $\mathcal{V}_U$ in terms of spin operators acting on the sites is~\cite{spinless1990}:
\begin{align}
H_{U} 
=& \tilde{H}_{RM} 
+ J_s \sum_i \mathbf{S}_{i,A} \cdot \mathbf{S}_{i,B} + J_w \sum_i \mathbf{S}_{i,B} \cdot \mathbf{S}_{i+1,A}
+ H_{\text{3-site}} + H_{\text{nn}} \,,
\end{align}
with $J_s = \frac{4 t_s^2\, U}{U^2 - \delta v^2} \approx 4t_s^2/U$ and $J_w = \frac{4 t_w^2\, U}{U^2 - \delta v^2} \approx 4t_w^2/U$ since $U$ is the largest scale. 
$\tilde{H}_{RM} $ is written in terms of fermions acting within $\mathcal{V}_U$: $\tilde{c}_{i\sigma} = P_U c_{i\sigma} P_U$, where $P_U$ is the projection operator onto $\mathcal{V}_U$. 
$H_{\text{nn}} = -J \sum_i n_{i,A} n_{i,B}/4 - J' \sum_i n_{i,B} n_{i+1,A}/4 $ is a density-density interaction term,
and $H_{\text{3-site}}$ is the three-site term \[H_{\text{3-site}} = -\frac{t_s t_w}{U} \sum_{i, \alpha, \beta} \left[ \left( \tilde{c}^{\dagger}_{i+1,A,\alpha} \v{\sigma}_{\alpha\beta} \tilde{c}_{i,A,\beta} \cdot \v{S}_{i,B} + \text{h.c.} \right) + (A \leftrightarrow B) \right].\]

To further obtain $H_{\text{low}}$ which is the spin model for the insulator at $1/4$ filling, we must systematically integrate out charge fluctuations. 
A simple example case is the limit with no hopping modulation, and only a sublattice potential $\delta v$. When $\delta v \gg t_s$, $\delta v$ sets the charge gap and we can obtain an effective spin model within one of the sublattices. The physical processes that involve the other sublattice can be eliminated to the desired order in $t_s/\delta v$ via a SW transformation. However, $t_s$ and $\delta v$ can be of comparable strength in our case. It is therefore important to generalize this approach by projecting onto the hybridized orbital basis to accurately capture the low-energy physics.

Let us split $H_{U}$ into intra and inter-unit cell terms.
\begin{align}
    H_U &= T_{\text{intra}} + T_{\text{inter}} + H_J + H_{\text{nn}} + H_{\text{3-site}}  \nonumber \\
    T_{\text{intra}} &= \sum_i -t_s \tilde{c}^{\dagger}_{i,A,\alpha} \tilde{c}_{i,B,\alpha} + h.c. + \delta v \(\tilde{c}^{\dagger}_{i,A,\alpha} \tilde{c}_{i,A,\alpha} - \tilde{c}^{\dagger}_{i,B,\alpha} \tilde{c}_{i,B,\alpha} \)\nonumber \\
    T_{\text{inter}} &= -t_w\sum_i \tilde{c}^{\dagger}_{i,B,\alpha} \tilde{c}_{i+1,A,\alpha} + h.c. \nonumber \\
    H_J &= \sum_i J \v S_{i,A} \cdot \v S_{i,B} + J'\v S_{i,B} \cdot \v S_{i+1,A}
\end{align}
The interaction term $H_{\text{nn}}$ and the three-site term $H_{\text{3-site}}$ do not affect the spin-couplings that we are after, and therefore, they are not explicitly written in the following.
Due to the hierarchy of energy scales, $T_{\text{intra}}$ is the leading term that we use to define the low-energy subspace $\mathcal{V}_{\Delta}$. Diagonalizing $T_{\text{intra}}$,
\begin{align}
    T_{\text{intra}} = \sum_{i,\alpha} \frac{\Delta_W}{2} \(\tilde{c}^{\dagger}_{i,+,\alpha} \tilde{c}_{i,+,\alpha} - \tilde{c}^{\dagger}_{i,-,\alpha} \tilde{c}_{i,-,\alpha}\)
\end{align}
where $+$ and $-$ label the two orbitals in each unit cell:
\begin{align}
    \tilde{c}^{\dagger}_{i,+,\alpha} &= \frac{\cos \frac{\theta}{2} \tilde{c}^{\dagger}_{i,A,\alpha} + \sin \frac{\theta}{2} \tilde{c}^{\dagger}_{i,B,\alpha}}{\sqrt{2}} \nonumber \\
    \tilde{c}^{\dagger}_{i,-,\alpha} &= \frac{-\sin \frac{\theta}{2} \tilde{c}^{\dagger}_{i,A,\alpha} + \cos \frac{\theta}{2} \tilde{c}^{\dagger}_{i,B,\alpha}}{\sqrt{2}} 
\end{align}
with $\tan \theta = t_s/\delta v$ and $\Delta_W = 2\sqrt{t_s^2 + \delta v^2}$. We will treat $T_{\text{inter}} + H_J$ perturbatively. 
The unperturbed ground state is then one in which the orbital labelled by $-$ is occupied in every unit cell. 
As in usual SW expansions, let us organize the subleading terms in the Hamiltonian by the number of orbital excitations they create. For notational convenience, we define
\begin{align}
    \v{S}_{i,ab} \equiv \frac{1}{2} \tilde{c}^{\dagger}_{i,a,\alpha} \v{\boldsymbol\sigma}_{\alpha \beta} \tilde{c}_{i,b,\beta}
\end{align}
where $a$ and $b$ are orbital labels.
Firstly there are terms that do not create any excitations:
\begin{align}
    \Pi_0 =& \Pi_{0,t_w} + \Pi_{0,J_s} + \Pi_{0,J_w}  \nonumber \\
    \Pi_{0,t_w} =& -\frac{t_w \sin \theta}{2} \sum_i \tilde{c}^{\dagger}_{i,+,\alpha} \tilde{c}_{i+1,+,\alpha} - \tilde{c}^{\dagger}_{i,-,\alpha} \tilde{c}_{i+1,-,\alpha} + h.c. \nonumber \\
    \Pi_{0,J_s} =& J_s \sum_i \(1 - 
    \frac{\sin^2 \theta}{2}\) \v S_{i,--} \cdot \v S_{i,++} - \frac{\sin^2 \theta}{4} \(\v S_{i,+-} \cdot \v S_{i,-+} + \v S_{i,-+} \cdot \v S_{i,+-} \) \nonumber \\
    \Pi_{0,J_w} =& J_w\sum_i \(\sin^2 \frac{\theta}{2} \v S_{i,++} + \cos^2 \frac{\theta}{2} \v S_{i,--}\) \cdot \(\cos^2 \frac{\theta}{2} \v S_{i+1,++} + \sin^2 \frac{\theta}{2} \v S_{i+1,--}\) \nonumber \\
    &- J_w\frac{\sin^2 \theta}{4} \(\v S_{i,+-} \cdot \v S_{i+1,-+} + \v S_{i,-+} \cdot \v S_{i+1,+-} \)
\end{align}
The terms that create one orbital excitations are:
\begin{align}
    \Pi_1 =& \Pi_{1,t_w} + \Pi_{1,J_s} + \Pi_{1,J_w} \nonumber \\
    \Pi_{1,t_w} =& -t_w \sum_i \cos^2 \frac{\theta}{2} \tilde{c}^{\dagger}_{i+1,+,\alpha} \tilde{c}_{i,-,\alpha} - \sin^2 \frac{\theta}{2} \tilde{c}^{\dagger}_{i,+,\alpha} \tilde{c}_{i+1,-,\alpha} \nonumber \\
    \Pi_{1,J_s} =& J_s \frac{\sin \theta}{2} \sum_i \(\cos^2 \frac{\theta}{2} \v S_{i,++} + \sin^2 \frac{\theta}{2} \v S_{i,--}\) \cdot \v S_{i,+-}   \nonumber \\
    &\qquad \qquad \qquad \qquad -\v S_{i,+-} \cdot \(\sin^2 \frac{\theta}{2} \v S_{i,++} + \cos^2 \frac{\theta}{2} \v S_{i,--}\) \nonumber \\
    \Pi_{1,J_w} =& -J_w \frac{\sin \theta}{2} \sum_i \(\sin^2 \frac{\theta}{2} \v S_{i,++} + \cos^2 \frac{\theta}{2} \v S_{i,--}\) \cdot \v S_{i+1,+-} \nonumber \\
    &\qquad \qquad \qquad \qquad - \v S_{i,+-} \cdot \(\cos^2 \frac{\theta}{2} \v S_{i+1,++} + \sin^2 \frac{\theta}{2} \v S_{i+1,--}\)
\end{align}
Terms that annihilate one orbital excitation are obtained by $\Pi_{-1}= \Pi_1^{\dagger}$. 
Additionally, there are terms that create two orbital excitations:
\begin{align}
    \Pi_{2} =& -J_w \frac{\sin^2 \theta}{4} \v S_{i,+-} \cdot \v S_{i+1,+-}
\end{align}

With the subleading contributions established, we proceed to the second stage of the SW transformation. Our objective is to decouple the low-energy manifold $\mathcal{V}_{\Delta}$ from the high-energy sectors by systematically eliminating the transition operators $\Pi_1$ and $\Pi_2$. This allows us to derive the effective Hamiltonian to a desired order $p$ in the small parameters $t_w$, $J_w$, and $J_s$ relative to the orbital gap $\Delta_W$: $\mathcal{O} \left( \frac{t_w^{p_1} J_w^{p_2} J_s^{p-p_1-p_2}}{\Delta_W^p} \right)$. As in standard $t/U$ expansions~\cite{macdonald1988}, this is enabled by the commutation relation:
\begin{align}
    [T_{\text{intra}}, \Pi_m] = m\Delta_W \Pi_m
\end{align}
It is important to note the validity of the commutation relation even when the electron operators $\tilde{c}$ are projected onto $\mathcal{V}_U$.
By identifying the specific terms within the SW series that contribute to the effective Hamiltonian, we establish the scaling of the spin couplings with respect to the parameters in the Hamiltonian $H_U$. 
These terms consist of products of $\Pi$ operators constrained by the requirement that system returns to the subspace $\mathcal{V}_{\Delta}$~\cite{macdonald1988}.
This is subsequently complemented by numerical extraction, which serves to both corroborate the predicted scaling laws and determine the precise magnitudes of the exchange constants. Accordingly, we focus on the operator scaling and do not explicitly track the algebraic coefficients in the following.

Our interest is in the low-energy spin Hamiltonian:
\begin{align}
H_{\text{low}} &= H_{\text{low}}^{(1)} + H_{\text{low}}^{(2)}, \\
H_{\text{low}}^{(1)} &= J_1 \sum_{i} \vec{S}_i \cdot \vec{S}_{i+1}, \qquad H_{\text{low}}^{(2)} = J_2 \sum_{i} \vec{S}_i \cdot \vec{S}_{i+2}\nonumber
\end{align}
We first derive the nearest-neighbor (NN) spin-coupling $J_1$. The leading contribution from the SW expansion is
\begin{align}
    H_{\text{low}}^{(1)} &=  P_{\Delta} \Pi_0 P_{\Delta} +  \frac{1}{\Delta_W^2} P_{\Delta}\Pi_{-1} \Pi_{0} \Pi_{1} P_{\Delta} \nonumber \\
    &= P_{\Delta} \Pi_0 P_{\Delta} +  \frac{1}{\Delta_W^2} P_{\Delta}\Pi_{-1,t_w} \Pi_{0,J_s} \Pi_{1,t_w} P_{\Delta} \nonumber \\
    &=\sum_i \frac{J_w \sin^2 \theta}{4}\v S_{i,--} \cdot \v S_{i+1,--}  \nonumber \\
    &\qquad \qquad + \frac{t_w^2 J_s}{\Delta_W^2} \cos^4 \frac{\theta}{2} \(1-\frac{\sin^2 \theta}{2}\) \tilde{c}^{\dagger}_{i,-,\alpha} \tilde{c}_{i+1,+,\alpha} \v S_{i+1,--} \cdot  \v S_{i+1,++} \tilde{c}^{\dagger}_{i+1,+,\beta} \tilde{c}_{i,-,\beta} \nonumber \\
    &= \mathcal{J}_1(\theta) \frac{t_w^2 J_s}{\Delta_W^2} \sum_i \v S_{i,--} \cdot \v S_{i+1,--} + \text{const.}
    \label{eq:app_J1}
\end{align}
where $\mathcal{J}_1$ is a dimensionless function of $\theta$. 
The first contribution is a consequence of the spatial delocalization of the orbitals within the unit cell; the lower orbital possesses a finite weight on both sublattices, allowing spins in adjacent cells to couple via the bare inter-cell spin coupling $J_w$. 
The second term represents a higher-order superexchange process. Unlike the direct term, this coupling involves the virtual hopping ($t_w$) of an electron into an adjacent unit cell. This transition creates a virtual orbital excitation during which the itinerant electron interacts with the neighboring spin via the intra-cell exchange $J_s$. Upon returning to its original cell, this process leaves behind an effective spin-spin interaction between the lower-orbitals of the two cells.

\begin{figure}[t]
\centering

\subfloat[$J_1(t)U$ as a function of $\omega t$ extracted\\from a fit of $J_1$ vs. $1/U$. \label{fig:app_J1_unscaled}]{%
  \includegraphics[width=0.5\columnwidth]{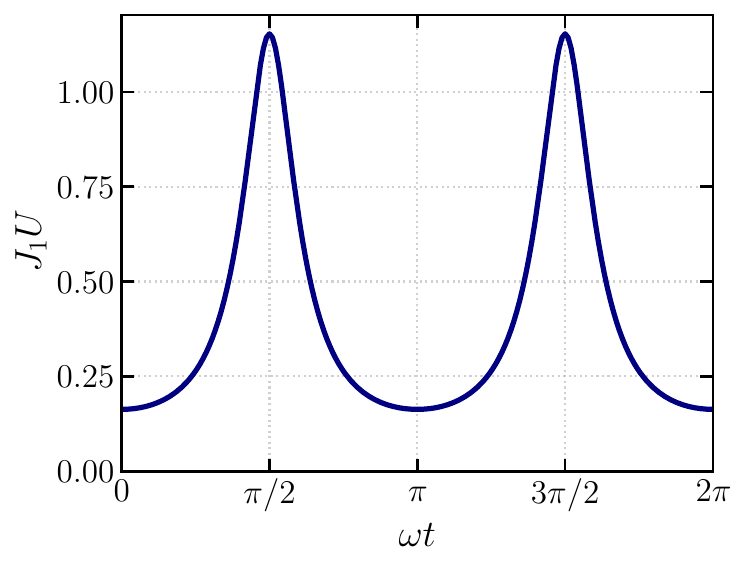}%
}
\hfill
\subfloat[$J_2(t)U^2$ as a function of $\omega t$ extracted\\from a fit of $J_2$ vs. $1/U^2$.\label{fig:app_J2_unscaled}]{%
  \includegraphics[width=0.49\columnwidth]{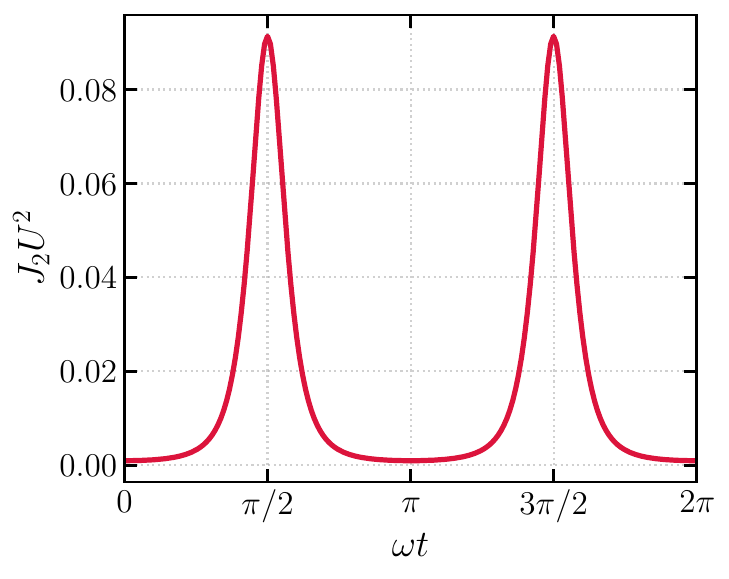}%
} 

\vspace{0.5ex} 

\subfloat[Dimensionless couplings $\mathcal{J}_1$ and $\mathcal{J}_2$ vs. $\omega t$.\label{fig:app_t}]{%
  \includegraphics[width=0.49\columnwidth]{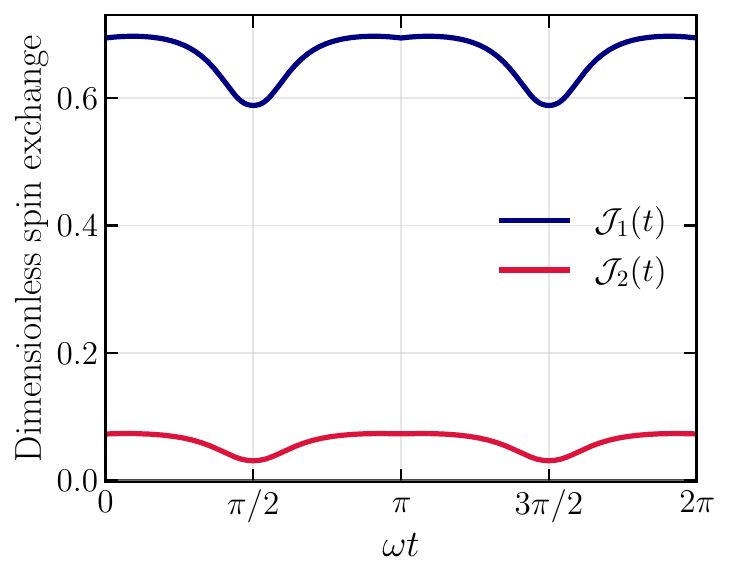}%
}
\hfill
\subfloat[Transformed coupling $\tilde{\mathcal{J}}_2 = \frac{\mathcal{J}_2(\tilde{t})\,\bar{J}_1}{J_1(\tilde{t})}$ vs. $\omega \tilde{t}$.\label{fig:app_tau}]{%
  \includegraphics[width=0.49\columnwidth]{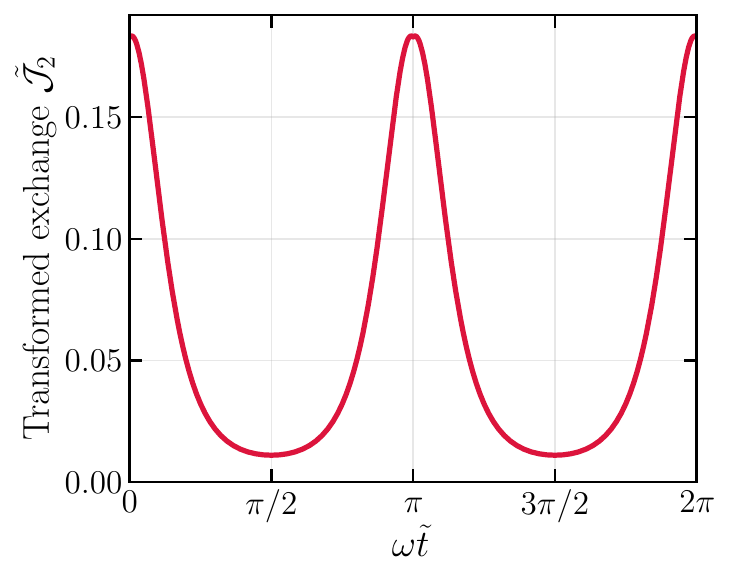}%
}
\caption{Spin exchange coefficients as a function of time $t$ and rescaled time $\tilde{t}$. Parameters used: $t_0 = 1$, $\delta t = 0.3$, $v_0 = 4$, and system size $L=6$ with $N=3$ particles. }
\label{fig:app_exchange}
\end{figure}

Turning now to the next-nearest neighbor (NNN) spin-coupling, the contributing terms within the Schrieffer-Wolff series expand as follows, where we have dropped constant contributions:
\begin{align}
    H_{\text{low}}^{(2)} = H_{1}^{(2)} + H_{2}^{(2)} + H_{3}^{(2)} + H_{4}^{(2)} + H_{5}^{(2)}
\end{align}
where
\begin{align}
H_1^{(2)} &\propto -\frac{1}{\Delta_W} P_{\Delta} \Pi_{-1,J_w} \Pi_{1,J_w} P_{\Delta}\nonumber \\
    &= \frac{J_w^2}{\Delta_W} \frac{\sin^2 \theta}{4} \sin^2 \frac{\theta}{2} \sum_i \left[ \left(\mathbf{S}_{i,--} \cdot \mathbf{S}_{i+1,-+}\right) \left(\mathbf{S}_{i+1,+-} \cdot \mathbf{S}_{i+2,--}\right) + \text{H.c.} \right] \nonumber \\
    &= \frac{J_w^2}{\Delta_W} \frac{\sin^2 \theta}{8} \sin^2 \frac{\theta}{2} \sum_i \left(\mathbf{S}_{i,--} \cdot \mathbf{S}_{i+2,--}\right) \\[2ex]
H_{2}^{(2)} &\propto  \frac{1}{\Delta_W^3} P_{\Delta}\Pi_{0,J_w} \Pi_{-1,t_w} \Pi_{0,J_s} \Pi_{1,t_w} P_{\Delta} \nonumber \\
    &\propto \frac{J_s J_w t_w^2}{\Delta_W^3} \sin^2 \theta \sum_i \left(\mathbf{S}_{i,--} \cdot \mathbf{S}_{i+1,--}\right) \left(\mathbf{S}_{i+1,--} \cdot \mathbf{S}_{i+2,--}\right) \nonumber \\
    &\propto \frac{J_s J_w t_w^2}{\Delta_W^3} \sin^2 \theta \sum_i \left(\mathbf{S}_{i,--} \cdot \mathbf{S}_{i+2,--}\right) 
    \end{align}
     \begin{align}
H_3^{(1)} &\propto \frac{1}{\Delta_W^3} P_{\Delta} \left(\Pi_{-1,t_w} \Pi_{0,J_s} \Pi_{0,t_w} \Pi_{1,J_w} + \Pi_{-1,J_w} \Pi_{0,t_w} \Pi_{0,J_s} \Pi_{1,t_w} \right) P_{\Delta} \nonumber \\
    &\propto \frac{J_s J_w t_w^2}{\Delta_W^3} \sin^2 \theta \sum_i \left[ \left(\mathbf{S}_{i,--} \cdot \mathbf{S}_{i+1,-+}\right) \left(\mathbf{S}_{i+1,+-} \cdot \mathbf{S}_{i+2,--}\right) + \text{H.c.} \right] \nonumber \\
    &\propto \frac{J_s J_w t_w^2}{\Delta_W^3} \sin^2 \theta \sum_i \left(\mathbf{S}_{i,--} \cdot \mathbf{S}_{i+2,--}\right) \\[2ex]
H_4^{(2)} &\propto \frac{1}{\Delta_W^3} P_{\Delta} \left(\Pi_{-1,J_s} \Pi_{0,t-w} \Pi_{0,J_w} \Pi_{1,t_w} + \Pi_{-1,t_w} \Pi_{0,J_w} \Pi_{0,t_w} \Pi_{1,J} \right) P_{\Delta} \nonumber \\
    &\propto \frac{J_s J_w t_w^2}{\Delta_W^3} \sin^2 \theta \sum_i \left[ \left(\mathbf{S}_{i,--} \cdot \mathbf{S}_{i+2,-+}\right) \left(\mathbf{S}_{i+2,+-} \cdot \mathbf{S}_{i+2,--}\right) + \text{H.c.} \right] \nonumber \\
    &\propto \frac{J_s J_w t_w^2}{\Delta_W^3} \sin^2 \theta \sum_i \left(\mathbf{S}_{i,--} \cdot \mathbf{S}_{i+2,--}\right) 
\end{align}
\begin{align}
H_5^{(2)} &\propto P_{\Delta} \frac{1}{\Delta_W^5}\Pi_{-1,t_w} \Pi_{0,J_s} \Pi_{1,t_w} \Pi_{-1,t_w} \Pi_{0,J_s} \Pi_{1,t_w} P_{\Delta} \nonumber \\
    &\propto \frac{J_s^2 t_w^4}{\Delta_W^5} \sum_i \left(\mathbf{S}_{i,--} \cdot \mathbf{S}_{i+1,--}\right) \left(\mathbf{S}_{i+1,--} \cdot \mathbf{S}_{i+2,--}\right) \nonumber \\
    &\propto \frac{J_s^2 t_w^4}{\Delta_W^5} \sum_i \left(\mathbf{S}_{i,--} \cdot \mathbf{S}_{i+2,--}\right)
\end{align}

Assembling all the contributions, we obtain the result:
\begin{align}
    H_{\text{low}}^{(2)} = \mathcal{J}_2(\theta) \frac{t_w^4}{\Delta_W^5} J_s^2 \sum_i \(\v{S}_{i,--} \cdot \v{S}_{i+2,--} \)
    \label{eq:app_J2}
\end{align}
where $\mathcal{J}_2(\theta)$ is a dimensionless function of $\theta$, which is finite in the limit $\theta \rightarrow 0$.

\begin{figure}[t]
    \centering
    \includegraphics[width=0.75\columnwidth]{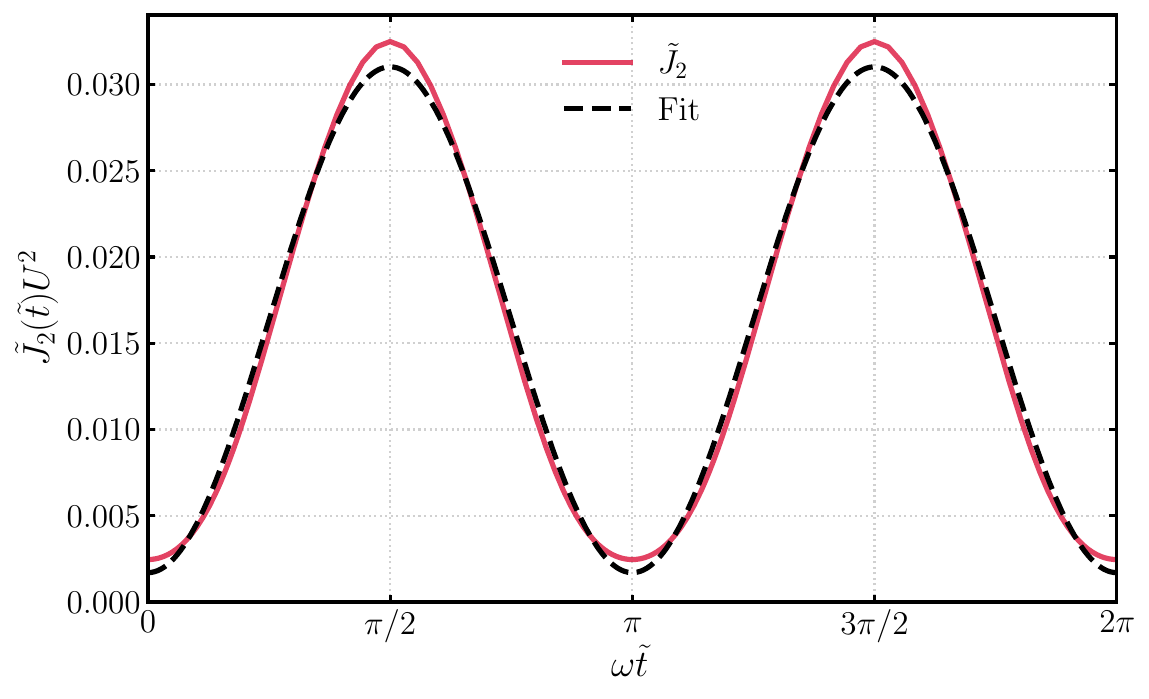}
    \caption{Transformed $\tilde{J}_2 U^2$ as a function of $\omega \tilde{t}$ (red). The dashed black line shows a first-harmonic fit, $\tilde{J}_2(\tilde{t}) = \bar{J}_2 + \delta J \cos(\omega \tilde{t})$, with $\bar{J}_2 \approx 0.016/U^2$ and $\delta J \approx -0.015/U^2$. Parameters: $t_0 = 1$, $\delta t = 0.3$, $v_0 = 4$, system size $L = 6$, and $N = 3$ particles.}
    \label{fig:harmonic_fit}
\end{figure}

For a three-spin cluster having six sites, we numerically extract  $J_1$ and $J_2$, and confirm the scaling results obtained in this section. 
We consider the fermionic system with $L=6$ sites with $N=3$ particles and open boundary conditions. In the Mott regime, this Hamiltonian is expected to have eight spin states as the lowest-energy eigenstates in the spectrum. Via Exact Diagonalization (ED), we obtain the lowest eight eigenstates. We project these eigenstates onto the subspace formed by the $-$ orbitals, defined earlier in this section, which well approximates the charge-wavefunction of the eigenstates. Within this subspace, we construct an effective Hamiltonian, which we then compare to the spin Hamiltonian for three spins having couplings $J_1$ and $J_2$. This allows us to extract $J_1$ and $J_2$ as a function of $U$,  which we use to confirm the scalings reported above. 
The extracted coefficients are plotted as a function of $\omega t$ in Fig.~\ref{fig:app_J1_unscaled} and Fig.~\ref{fig:app_J2_unscaled}.
Further, we obtain the dimensionless couplings $\mathcal{J}_1$ and $\mathcal{J}_2$ as a function of time for our drive protocol. 
Figures~\ref{fig:app_t} and \ref{fig:app_tau} show the evolution of $\mathcal{J}_1$ and $\mathcal{J}_2$ over a single cycle of the pump. 

Using the conventions introduced in the main text, we obtain, for the next-nearest-neighbor coupling at $t_0 = 1$, $\delta t = 0.3$, and $\delta v = 4$, $\bar J_2 + \delta J \cos(\omega \tilde{t})$, with $\bar J_2 \approx 0.016/U^2$ and $\delta J \approx -0.015/U^2$. 
The transformed exchange $\tilde{J}_2$ is obtained from $\tilde{J}_2(\tilde{t}) = \bar{J}_1 J_2(\tilde{t})/J_1(\tilde{t})$ and is fitted to its first harmonic. $\tilde{J}_2(\tilde{t})$ together with the harmonic fit are shown in Fig.~\ref{fig:harmonic_fit}.

\section{Exact Diagonalization and Real-Time Evolution}
\label{app:numerics}

\subsection{Quarter-filling}
In this section, we provide details of the numerical calculations, performed using the QuSpin package~\cite{weinberg2017,weinberg2019}, for the interacting Rice-Mele model of Eq.~\eqref{eq:H1} at quarter filling, along with additional results.
We perform ED within the $S_z=0$ sector and the zero-momentum sector to obtain the low-lying states in the spectrum. Throughout, we set $t_0 = 1$, $\delta t = 0.5$, $\delta v = 1$ and choose a $1d$ chain with $L=16$ sites. Periodic boundary conditions were used throughout our calculations. Quarter filling and $S_z=0$ constrain the number of particles in each spin sector: $N_{\uparrow} = N_{\downarrow} = L/4$. In this symmetry sector, there are ten spin states. Since we are in the Mott regime, the ten lowest energy eigenstates correspond to these spin states. Any higher energy excitation necessarily involves the conduction band and thus adds particle-hole excitations in the charge sector on top of the ground state.

\begin{figure}[t]
\centering

\subfloat[$U=14.29$.
\label{fig:pumped_charge_vs_time_a}]{
  \includegraphics[width=0.48\columnwidth]{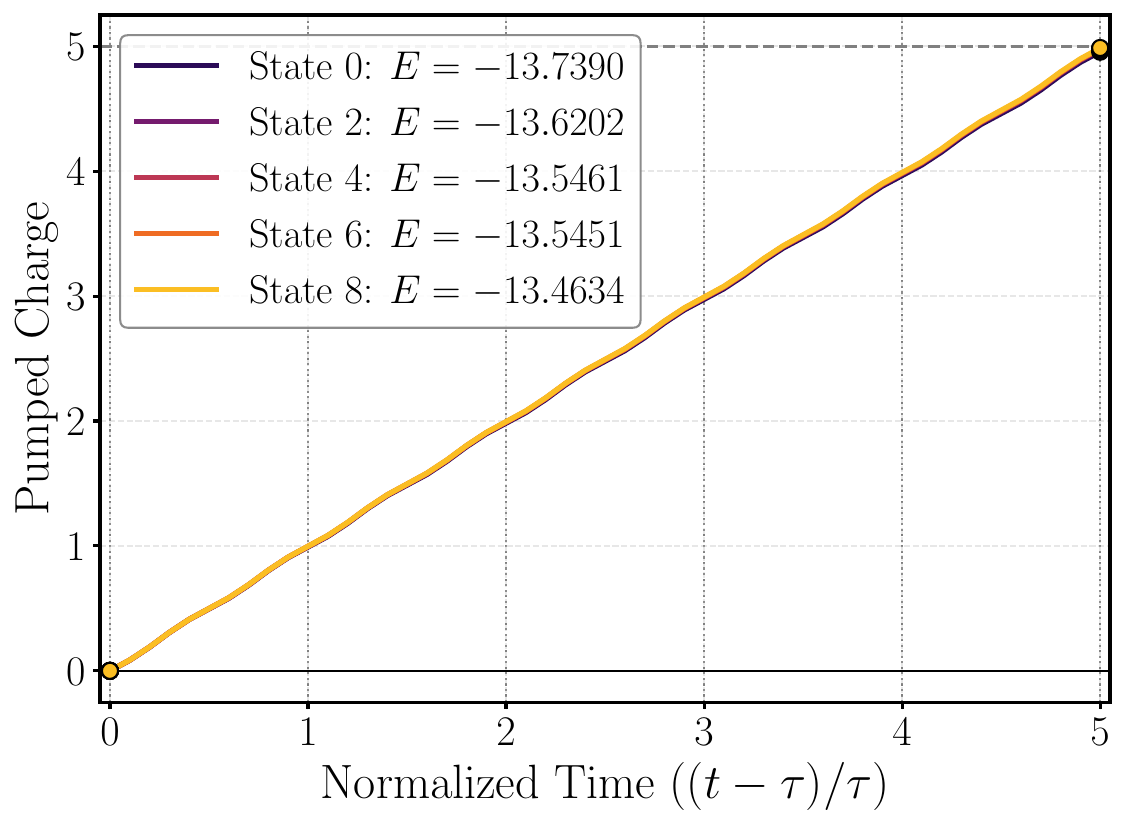}
}
\hfill
\subfloat[$U=3.33$.
\label{fig:pumped_charge_vs_time_b}]{
  \includegraphics[width=0.48\columnwidth]{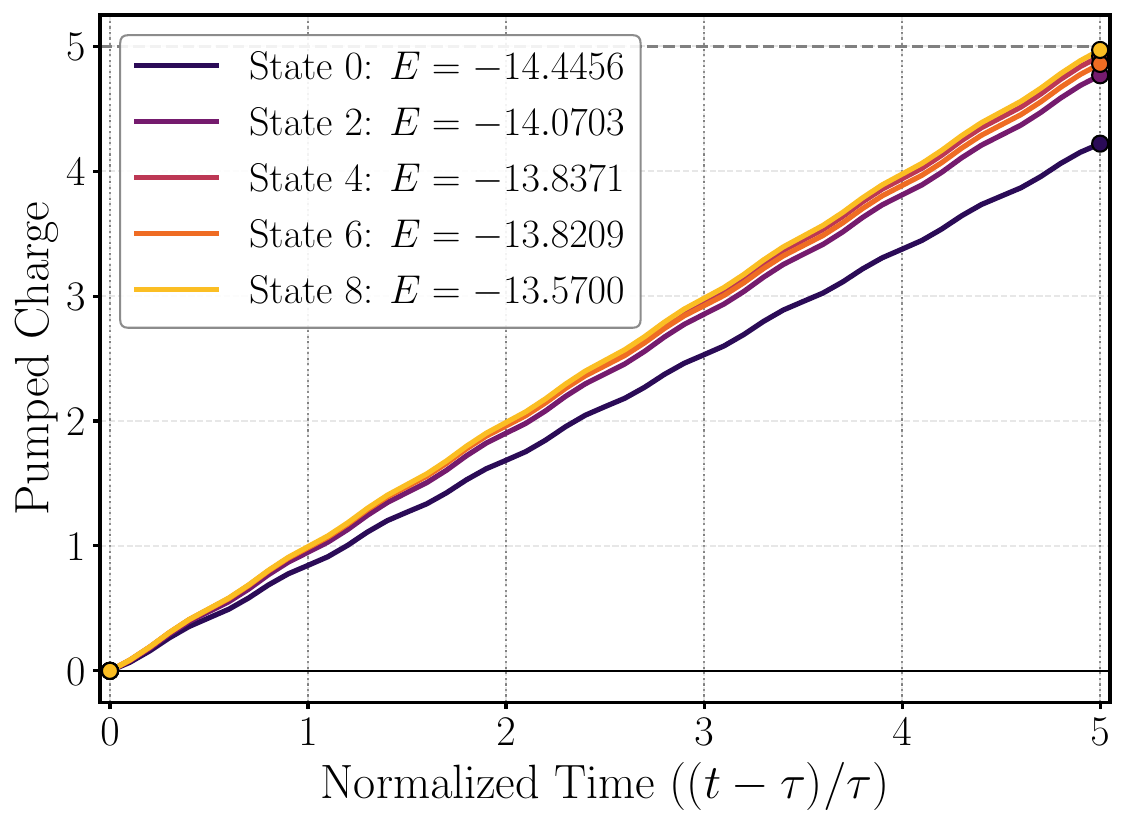}
}\caption{Pumped charge as a function of time for drive frequency $\omega = 0.1$ and two values of the on-site Hubbard interaction $U$. Curves are labeled by the index and energy eigenvalue $E$ of the corresponding initial many-body state within the energy-ordered spectrum of the chosen symmetry sector. Here, all the initial states are low-energy spin states. Parameters: $t_0 = 1$, $\delta t = 0.5$, $\delta v = 1$, and $L = 16$ sites.}
\label{fig:pumped_charge_vs_time}
\end{figure}

\begin{figure}[t]
    \centering
    \includegraphics[width=0.65\columnwidth]{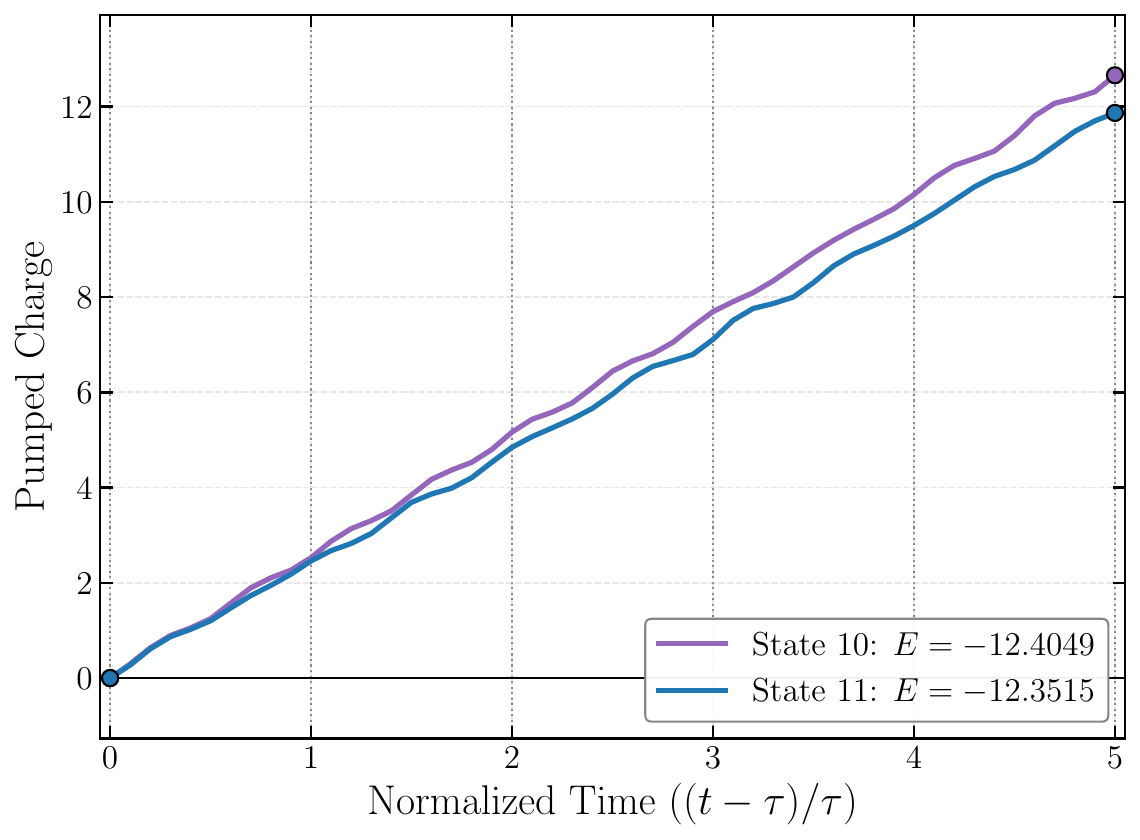}
    \caption{Pumped charge versus time for initial states with particle-hole excitations in the charge sector, for drive frequency $\omega = 0.1$ and on-site Hubbard interaction $U=8.26$. Curves are labeled by their corresponding initial many-body state index within the energy-ordered many-body spectrum of the chosen symmetry sector and energy eigenvalue $E$. }
    \label{fig:pumped_charge_vs_time_ph}
\end{figure}

To ensure adiabaticity during the time-periodic protocol in Eq.~\eqref{path1}, it is important to turn on the drive in a smooth manner. To this end, we implement a ramp protocol for a time interval $\tau$ during which we slowly increase the drive frequency. Namely, we implement the following instantaneous phase in the drive Hamiltonian:
\begin{align}
\phi(t) = 
\begin{cases} 
\omega \tau \left[ 70 \left(\frac{t}{\tau}\right)^5 - 224 \left(\frac{t}{\tau}\right)^6 + 280 \left(\frac{t}{\tau}\right)^7 - 160 \left(\frac{t}{\tau}\right)^8 + 35 \left(\frac{t}{\tau}\right)^9 \right], & 0 \leq t \leq \tau \\
\omega t, & t > \tau
\end{cases}
\end{align}
with $\delta t(t)=\delta t \sin\phi(t)$, $\delta v(t)=\delta v \cos \phi(t)$, which are parameters in the interacting Rice-Mele model defined in Eq.~\ref{path1}.
Importantly, this protocol ensures that the phase satisfies $\phi(0)=0$ and $\phi(\tau)=\omega \tau$, together with the smoothness conditions $\phi'(0)=0$ and $\phi'(\tau)=\omega$.
While these conditions are satisfied by lower-order polynomials in $t$, our choice results in a large suppression of the high-frequency ($\Omega \gg \omega$) tail in the Fourier transform of $\text{exp}(i\phi(t))$: $\sim 1/\Omega^6$. This is a consequence of $\phi''(t)=\phi'''(t) = \phi''''(t)=0$ at $t=0$ and $t=\tau$. In particular, for $\Omega \sim \Delta_c \gg \omega$, this protocol ensures a strong suppression of all effects from starting the ramp protocol.

We choose one of the low-lying eigenstates of the Hamiltonian at $t=0$ to be the initial state, and time-evolve it within the symmetry sector using the drive protocol discussed above. 
The time-evolved state is obtained via adaptive Runge-Kutta time integration using the QuSpin package~\cite{weinberg2017, weinberg2019}, from which we compute the expectation value of the current operator averaged over all the bonds $J(t)$.
The pumped charge is then given by $\Delta Q(t) = \int_0^t dt' J(t')$. 
Fig.~\ref{fig:pumped_charge_vs_time} displays the numerical results for the time-dependent pumped charge, $\Delta Q(t)$. Here, the pumped charge is plotted relative to the state at $t = \tau$.
In Fig.~\ref{fig:pumped_charge_vs_time_a}, we track the time-dependent pumped charge, $\Delta Q(t)$, initialized from various low-energy spin configurations. Consistent with our analytical expectations, the pump displays robust quantization deep within the Mott-insulating regime, with the accumulated charge per cycle closely tracking the integer topological value, $\Delta Q = 1$. 
The observed breakdown of exact quantization at longer times is associated with drive-induced heating. 

As shown in Fig.~\ref{fig:pumped_charge_vs_time}, this heating-induced deterioration is most prominent for the system initialized in its ground state. As discussed  at the end of Sec.~\ref{sec:heatingCharge}, the counterintuitive behavior is explained by the fact that the highest-energy state is close to a ferromagnetic state. Fully spin-polarized electrons are non-interacting within our Hubbard model and thus inelastic scattering is strongly suppressed for high-energy states in the spin sector. Note that in a high-temperature state of a large system, such states carry an exponentially small weight and can safely be ignored. 
As discussed in the main text, quantization of the topologically pumped charge improves systematically with increasing~$U$.

We show numerical results for the sector with $S_z=0$ and total momentum $P=0$.
 Here, one has to take into account that a fully filled band of spinless fermions carries the total momentum $P=\pi/a$. That implies, that to avoid charge excitations, our spin system is forced into a sector with total momentum $\pi/a$. Therefore, $S_z=0$ sector does contain a fully polarized state (obtained by applying $(\sum_i S^-_i)^{N/2}$ to the spin-up polarized states).
Starting from a ferromagnetic state, the finite-momentum spin-state can be realized by creating $O(M)$ magnons with momenta 
$O(\pi/(aM))$ each. This lowers the energy of the ferromagnetic state only by  $\sim  -J M/M^2\sim -J/M$. Consistent with numerics, in such a state inelastic processes are strongly suppressed (this can be viewed as a consequence of the Goldstone theorem).

Starting from initial states that possess excitations across the charge gap, see Fig.~\ref{fig:pumped_charge_vs_time_ph}, the pumped charge is not quantized, not universal, differs from cycle to cycle and is strongly affected by finite-size effects.

\begin{figure}[t]
    \centering
    \includegraphics[width=0.9\columnwidth]{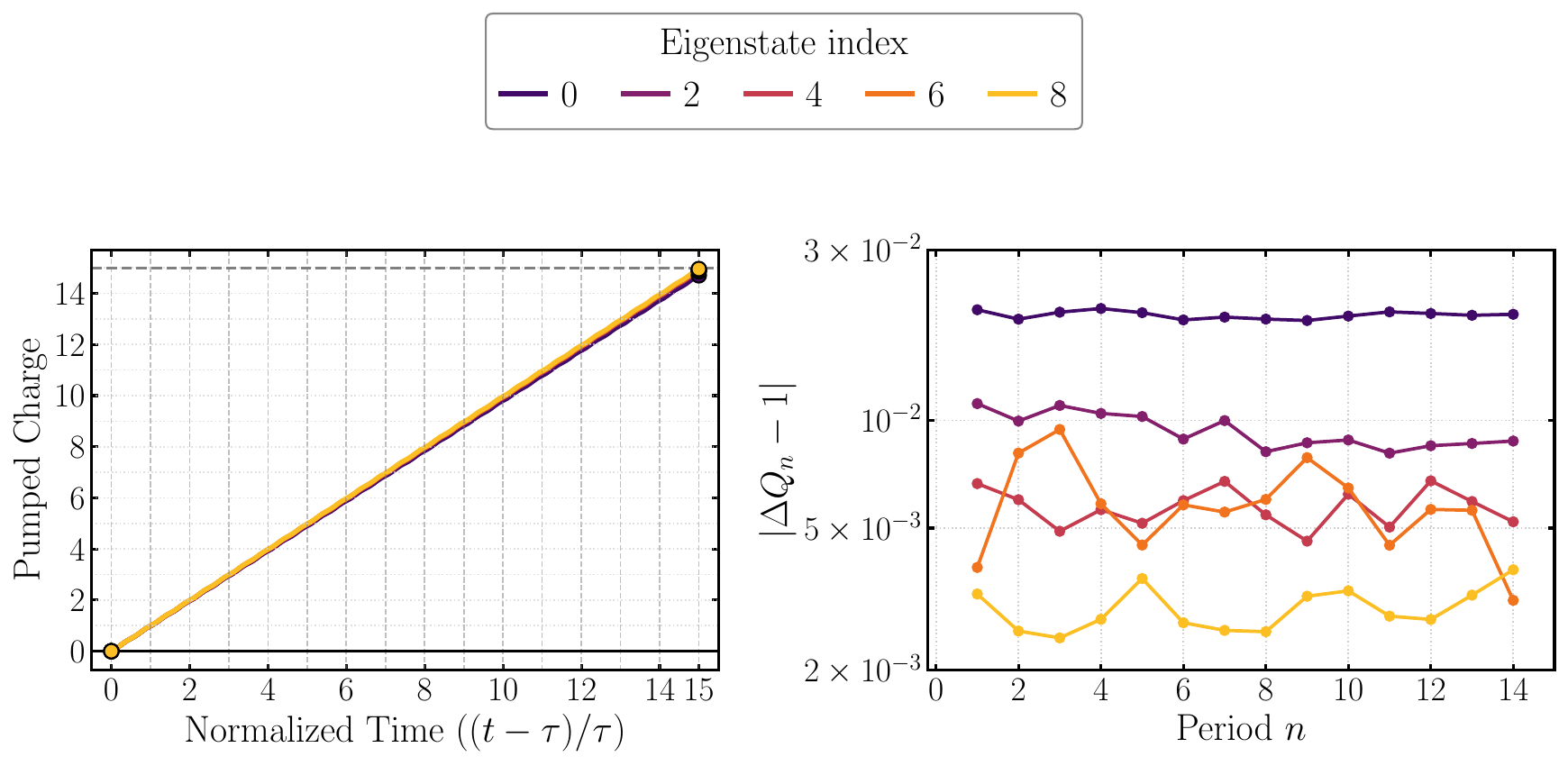}
    \caption{Left panel: Pumped charge as a function of time over $15$ drive periods for a large interaction strength $U$, showing robust topological charge pumping. Right panel: Deviation from perfect quantization of $\Delta Q_n$, the charge pumped during the $n$-th driving period, i.e., from time $n\tau$ to $(n+1)\tau$, as a function of $n$. Parameters: $U = 8.26$, $\omega = 0.2$, $L = 16$.}
    \label{fig:pumped_charge_vs_time_longtime}
\end{figure}

\begin{figure}[t]
    \centering
    \includegraphics[width=0.7\columnwidth]{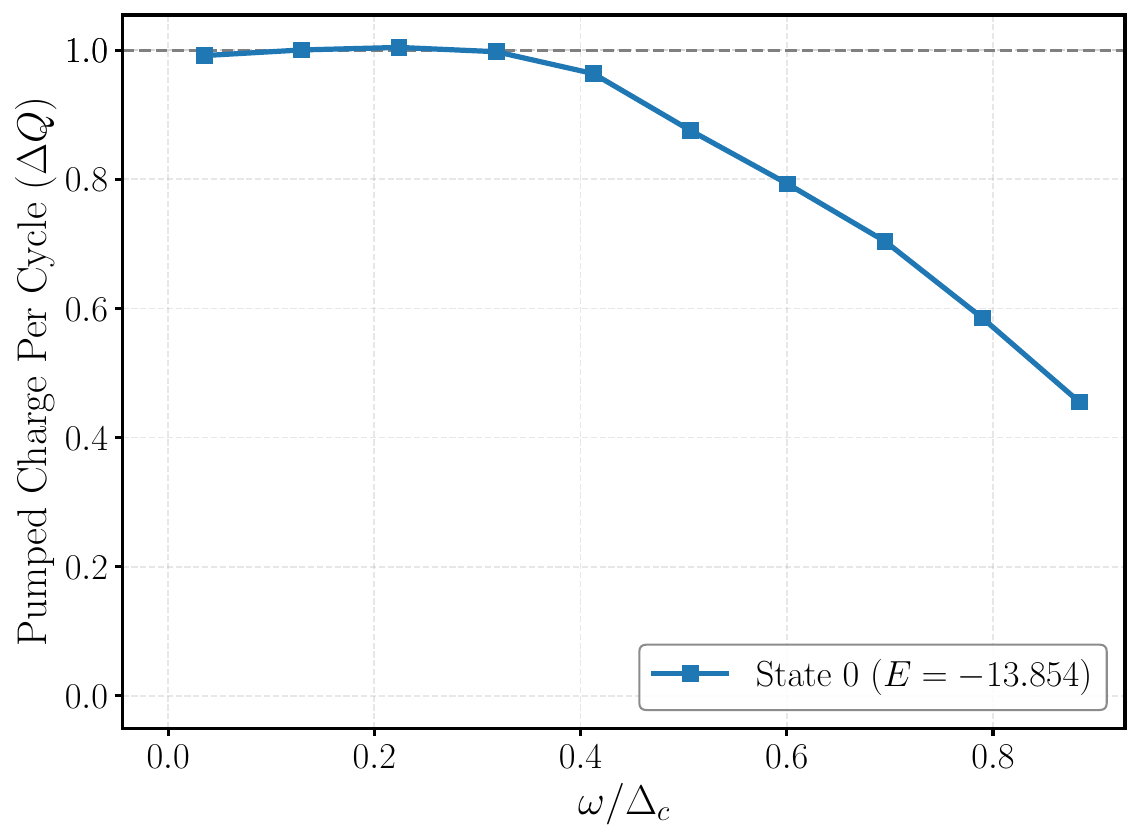}
    \caption{Pumped charge averaged over five cycles, plotted versus $\omega/\Delta_c$ for $U=10$. Time evolution is performed with the system prepared in the ground state at $t = 0$ for $L=16$.}
    \label{fig:pumped_charge_vs_omega}
\end{figure}

In Fig.~\ref{fig:pumped_charge_vs_omega} we show how the pumped charge depends on the driving frequency. While quantized pumping breaks down for large frequencies, it is remarkably robust at low frequencies, $\omega \lesssim 0.3\,\Delta_c$, where $\Delta_c = 2\sqrt{\delta v^2 + 4 \delta t^2}$ is the charge gap. This is consistent with an exponential protection of adiabatic pumping for small frequencies. 

\subsection{Half-filling}
\begin{figure}[t]
    \centering
    \includegraphics[width=0.65\columnwidth]{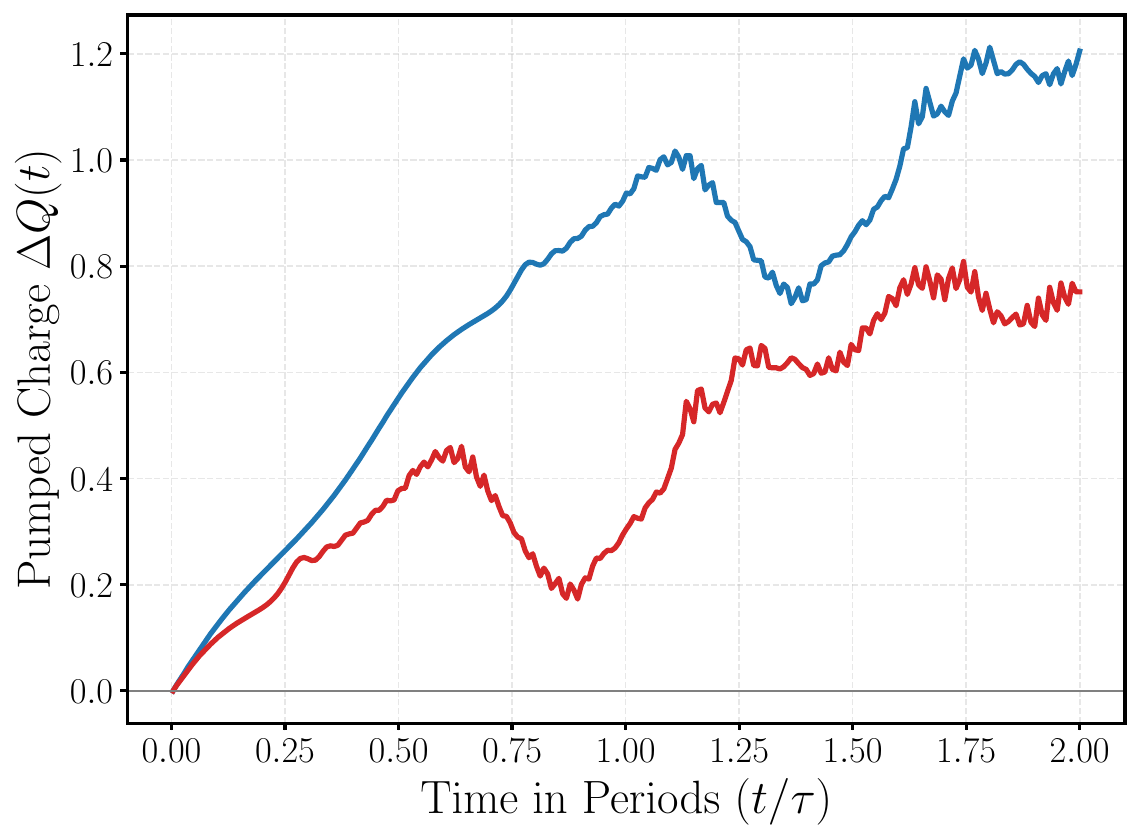}
    \caption{Pumped charge versus time at half-filling for the ground state. The blue/red curves correspond to the drive protocols defined by $\phi_{\pm}(t)$ in Eq.~\ref{eq:ramp_half}.  }
    \label{fig:half_filling}
\end{figure}

For the drive in Eq.~\eqref{eq:drive_half} of the main text at half-filling, we have shown that the pump is not expected to be robust. The main culprit being the closing of the spin gap during the pumping cycle and the conversion of spin-excitations into charge excitations. This case was also studied experimentally and numerically in Ref.~\cite{Viebahn2024}. In this section, we report our numerical results for the half-filling case obtained via ED. We use the parameters: $U=7.1$, $t_0 = 1$, $\delta t = 0.88$, $\delta v = 2.1$~\cite{Viebahn2024}, drive frequency $\omega = 2\pi/50$, number of sites $L=8$, with $N_{\uparrow} = N_{\downarrow} = 4$. Following Ref.~\cite{Viebahn2024}, we adopt the ramp protocols:
\begin{equation}
\phi_{\pm}(t)=
\begin{cases}
\dfrac{\omega}{3T_{\mathrm{ramp}}^2}
\left(t+T_{\mathrm{ramp}}\right)^3 \pm \pi/2,
& -T_{\text{ramp}} \le t \le 0,\\[6pt]
\omega t \pm \pi/2,
& t > 0.
\end{cases}
\label{eq:ramp_half}
\end{equation}
where $\phi_{\pm}(t)$ defines two pump protocols. The duration of the ramp $T_{\text{ramp}}$ is set to $T_{\text{ramp}} = \tau/16$. 
While the ramp ensures adiabaticity in the charge sector, the system undergoes a closing of the spin gap at $t = \frac{\tau}{2} \pm n\frac{\tau}{4}$, with $n$ being an integer, for the two protocols. 
We start with the ground state of the Hamiltonian at $t=-T_{\text{ramp}}$ and time-evolve it to compute the pumped charge, as done for the quarter-filling case. 
Our results, shown in Fig.~\ref{fig:half_filling} for $t>0$, agree qualitatively with results of Ref.~\cite{Viebahn2024} for $L=10$, demonstrating that the pump is destabilized by the closing of the spin gap.
For the protocol defined by $\phi_+$, the topological charge pumping survives up to $t \approx \tau$, whereas, the other protocol is more fragile due to the closing of the spin gap at an earlier time. 
These results reveal a striking contrast between the quarter- and half-filled cases, highlighting the unique suitability of the quarter-filled system for supporting robust topological charge pumping in the incoherent Mott regime.

\end{appendix}
\end{document}